\documentclass[preprint,12pt]{article}

\usepackage{authblk}
\usepackage[english]{babel}	
\usepackage[utf8x]{inputenc}		
\usepackage{enumerate}
\usepackage{textcomp}                
\usepackage{typearea}
\usepackage{pdfpages}
\usepackage[toc]{appendix}
\usepackage[]{todonotes}
\usepackage{subcaption}

\usepackage{comment}
\usepackage{afterpage}
\usepackage{placeins}

\usepackage{amsfonts, amsmath, amssymb, amsthm, dsfont}
\usepackage{amsthm}			
\usepackage{thumbpdf}	
\usepackage{natbib}

\usepackage{floatpag}
\usepackage{pgf, tikz}
\usepackage{tikz-cd}		
\usepackage{bbm}
\usepackage{bm}
\usepackage{algorithm2e}
\usepackage{algorithmic}

\theoremstyle{definition}

\newtheorem*{assumption*}{Assumption}

\newtheorem*{condition*}{Condition}

\theoremstyle{plain}

\theoremstyle{remark}

\usepackage{graphicx}
\usepackage{hyperref}

\newcommand{\R}{\mathbb{R}}

\newcommand{\E}{\mathbb{E}}

\newcommand{\F}{\mathcal{F}}

\newcommand{\Var}{\operatorname{Var}}

\newcommand{\vx}{\mathbf{x}}
\newcommand{\vv}{\mathbf{v}}
\newcommand{\vu}{\mathbf{u}}
\newcommand{\vz}{\mathbf{z}}
\newcommand{\vc}{\mathbf{c}}
\newcommand{\mSigma}{\mathbf{\Sigma}}

\newcommand{\vF}{\mathbf{F}}
\newcommand{\vW}{\mathbf{W}}
\newcommand{\vw}{\mathbf{w}}
\newcommand{\vmu}{\boldsymbol{\mu}}

\title{An efficient jump-diffusion approximation of the Boltzmann equation
}
\author[1]{Fabian Mies}
\author[2,3]{Mohsen Sadr}
\author[4]{Manuel Torrilhon}
\affil[1]{Institute of Statistics, RWTH Aachen University, Aachen, Germany}
\affil[2]{Swiss Plasma Center, EPFL, CH-1015 Lausanne, Switzerland
}
\affil[3]{Department of Mechanical Engineering, MIT, Cambridge, MA 02139, USA}
\affil[4]{MATHCCES, Department of Mathematics, RWTH Aachen University, Schinkelstrasse 2, Aachen, Germany}
\usepackage{cancel}
\begin{document}
	
\maketitle
\begin{abstract}
\noindent A jump-diffusion process along with a particle scheme is devised as an accurate and efficient particle 
solution to the Boltzmann equation. 
The proposed
process (hereafter Gamma-Boltzmann model) 
is devised to
match the evolution of all moments up to the heat fluxes 
while attaining
the correct Prandtl number of $2/3$ for monatomic gas 
with Maxwellian molecular potential. 
This approximation model is not subject to issues associated with the previously developed Fokker-Planck (FP) based models; such as having wrong Prandtl number, limited applicability, or requiring estimation of higher-order moments.
An efficient particle solution to  the proposed Gamma-Boltzmann model is 
devised
and 
compared computationally to the direct simulation Monte Carlo and the cubic FP model
[M. H. Gorji, M. Torrilhon, and P. Jenny, J. Fluid Mech. 680 (2011): 574-601] in several test cases including Couette flow and lid-driven cavity.
The simulation results indicate that the Gamma-Boltzmann model yields a good approximation of the Boltzmann equation, provides a more accurate solution compared to the cubic FP in the limit of a low number of particles, and remains computationally feasible even in dense regimes.

\end{abstract}
\section{Introduction}


\noindent As fluid flows depart from equilibrium, the underlying closure assumptions in the classical continuum description break down, see e.g. \cite{wang2003predicting}. In order to capture the physics of the non-equilibrium phenomena, a mathematical model from the smaller scale, i.e. mesoscale, needs to be considered. 
Kinetic theory provides an accurate statistical description of non-equilibrium fluid flows by introducing an evolution equation for the particle velocity distribution function. In the case of monatomic and neutral particles, assuming molecular chaos, and in the limit of low density, Boltzmann devised an exact evolution equation for the single-particle distribution function that undergoes a binary collision operator, see \cite{Chapman1953}.
\\ \ \\
Several approaches for solving the Boltzmann equation numerically have been developed in the literature, in particular the discrete velocity method, moment methods, and particle Monte Carlo algorithms among others. \\ \ \\
The various variants of the discrete velocity method discretize the phase space directly to solve a finite system of equations. While this approach provides accurate solutions, the computational cost limits this approach in practice, see e.g. \cite{broadwell_1964,platkowski1988discrete}. The major obstacle is the high dimensionality of the phase space and cutting off the velocity space.
The high dimensionality of the solution can be resolved by so-called moment methods, where finitely many moments of the particle distribution are considered, their evolution equations are derived from the Boltzmann operator, and the resulting system of partial differential equations is solved numerically \cite{Torrilhon2013,torrilhon2016modeling}. Although the moment methods allow for fast solutions, they require an
ansatz for the velocity distribution function, e.g., Grad's ansatz. This closure problem occurs because the evolution of moments typically depends on other moments of the distribution which are not solved for. Moreover, numerical challenges in incorporating boundary conditions and restriction due to the stability of the outcome moment system are introduced, see, e.g.,, \cite{torrilhon2016modeling,sarna2018stable}. 
\\ \ \\
A mature approach in solving the Boltzmann equation  consistently yet subject to the statistical noise is the direct simulation Monte Carlo (DSMC) as proposed by \cite{Bird1970}, see also \cite{Bird1994}. Here, the distribution in phase space is represented by a finite number of computational particles. These particles evolve according to the dynamics underlying the Boltzmann equation, and pairwise collisions are performed explicitly. Spatial heterogeneity is incorporated by splitting the domain into computational cells, performing collisions in each cell independent of others, and streaming the position of particles after the collision step successively. As the number of computational particles tends to infinity, Bird's method is expected to converge to the solution of the Boltzmann equation, see \cite{Myong2019} for a computational analysis.  A major theoretical result concerning the latter validity has been obtained by \cite{Wagner1992}, showing that the limiting distribution satisfies an equation which closely resembles Boltzmann's equation. However, \cite{Wagner1992} does not consider the limit as the space resolution increases. For an alternative simulation method proposed by \cite{Nanbu1980} (see also \cite{Nanbu1983}), the consistency is demonstrated rigorously by \cite{Babovsky1989} who also accounts for spatio-temporal discretization errors. A third approach is presented by \cite{Lukshin1988} for the spatially homogeneous setting. 
Consistency with the homogeneous Boltzmann equation is shown as the number of computational particles increases. 
While the DSMC method has been historically deemed to be prohibitively expensive, recent advances in parallel computing have increased its practical applicability, see \cite{Goldsworthy2014} and \cite{Plimpton2019}, among others. 
As the direct simulation methods need to resolve all binary collisions, they are computationally expensive at low Knudsen number regimes, i.e.\ where collisions becomes the dominant process.
\\ \ \\
Various attempts have been made to approximate the Boltzmann equation with a simpler model which provides a reasonable estimation of moments up to heat flux while allowing for improved numerics. In particular, a Fokker-Planck model with linear drift was devised as an efficient approximation to the Boltzmann equation \cite{Jenny2010}. The Fokker-Planck model may be related to particle dynamics driven by stochastic differential equations, and hence allows for a solution via particle Monte Carlo methods. In contrast to the original Boltzmann equation, the collisional jump process is replaced with a continuous movement which represents the collisions in an aggregated fashion. Hence, the resulting particle scheme can be extended to the dense regime (low Knudsen number) without introducing further numerical cost.
However, the linear Fokker-Planck model suffers from having a wrong Prandtl number. In order to resolve this issue, two main approaches have been suggested. First, the correct Prandtl number in the Fokker-Planck model was obtained by introducing a cubic drift and choosing the free parameters such that the relaxation rates of stress tensor and heat fluxes are consistent with the ones of Boltzmann equation for Maxwell molecules in the homogeneous setting \cite{Gorji2011}. Unfortunately, evaluating the projected coefficients in the cubic FP relies on estimation of moments up to fifth order from the particles which is 
prone to higher error in noisy scenarios than linear FP, since  the statistical error typically increases with the moment order. In the second approach devised by \cite{Mathiaud2016}, a linear FP model with a non-isotropic (ellipsoidal) diffusion tensor was devised to correct the Prandtl number. Compared to the cubic FP model, the drift remains linear in the drift, which yields computational advantages. However, the positive-definiteness of the non-isotropic diffusion tensor can no longer be guaranteed, such that this method is not applicable in all cases. 
\\ \ \\
In this paper, we present a new approach to fix the Prandtl number of linear FP model by introducing additive jumps, such that the trajectory of particles is governed by a jump-diffusion process. We suggest the Gamma process as a model for the jumps, and we choose the parameters carefully to match the correct relaxation rates of stress tensor and heat fluxes. We refer to this proposed model as the Gamma-Boltzmann model. In contrast to the cubic FP model of \cite{Gorji2011}, we only require estimates of moments up to third order, which is expected to improve the solution compared to cubic FP in noisy scenarios. In contrast to the ellipsoidal FP model of \cite{Mathiaud2016}, the diffusion tensor remains positive-semidefinite, such that our model is applicable in all situations. In contrast to the collisional jumps of the DSMC method, this jump-diffusion model aggregates the collisions and allows for fast simulation of the particle trajectories, even in the dense regime.
\\ \ \\
The remainder of this paper is structured as follows. In \S~\ref{sec:Rev_Boltzmann}, the Boltzmann equation and its Fokker-Planck approximation are reviewed. Next, the generic jump-diffusion model is presented in \S~\ref{sec:jump_diffusion}. We highlight that the exact Boltzmann equation may also be regarded as a specific jump-diffusion (\S~\ref{sec:Bolt_as_jump}), and we devise the Gamma-Boltzmann model with correct Prandtl number (\S~\ref{sec:Gamma-Boltzmann}). The corresponding particle Monte Carlo scheme is described in \S~\ref{sec:sampling}. In \S~\ref{sec:comp_res}, the solution obtained from the Gamma-Boltzmann model is tested against solution obtained from cubic FP model as well as DSMC for the Couette flow and the lid-driven cavity. Finally, in \S~\ref{sec:conclusion}, the conclusion and outlook for future works are provided.  
\\ \ \\
 In appendices A-C, technical derivations of the proposed Gamma-Boltzmann model are carried out. Furthermore, a digital supplement providing a brief but rigorous primer on jump-diffusion processes is attached to this manuscript for the reader.

\section{Review of the kinetic models}
\label{sec:Rev_Boltzmann}
\subsection{Kinetic theory and Boltzmann equation}

The state of a dilute, monatomic gas may be described via its velocity distribution at location $\vx\in\R^3$ and time $t\in\R^+$.
A convenient way to identify this distribution is in terms of its phase-space density $\F=\F(\vv, \vx,t)$, which represents the mass-weighted number of particles at time $t$ whose locations and velocity fall inside an infinitesimal volume around $(\vx,\vv)\in\R^6$.   
The particle distribution evolves via advection, external force field $\bm F$ and collisions between particles $\mathcal{S}\F$, i.e.,
\begin{align}
	\frac{\partial \F}{\partial t} + \sum_i \frac{\partial}{\partial x_i} (v_i \F) + \sum_i \frac{\partial}{\partial v_i} ( F_i \F )& = \mathcal{S}\F. \label{eqn:kinetic}
\end{align}
In particular, the Boltzmann collision operator, which only acts on the velocity $\vv$, takes the form 
\citep{Bird1994}
\begin{align*}
	\mathcal{S}^\text{Boltz} \F(\vv) &= \frac{1}{m} \int_{\R^3} \int_{0}^{4\pi} 
	\left[\F(\vv')\F(\bar{\vv}') - \F(\vv)\F(\bar{\vv})\right] \, |\vv-\bar{\vv}| I(\Omega, |\vv-\bar{\vv}|)\, d\Omega 
	\, d\bar{\vv}.
\end{align*}
Here, $m$ is the mass of a single particle, $(\vv', \bar{\vv}')$ are the post-collision velocities corresponding to a collision pair $(\vv, \bar{\vv})$,
$\Omega$ is the solid angle about the vector $\vv'- \bar{\vv}'$, and $I(\Omega, |\vv-\bar{\vv}|)d\Omega$ is the differential cross-section of the collision, see \cite{Bird1994} for details.
The exact form of $I$ depends on the specific molecular potential.
In this paper, we focus on Maxwellian molecules where  $|\vv-\bar{\vv}| I(\Omega, |\vv-\bar{\vv}|) d\Omega $ becomes independent of the relative velocity $|\vv-\bar{\vv}|$ which simplifies computation of moments, see \cite[2.8]{Bird1994} and \cite[5.3.3]{Struchtrup2005}.
\\ \ \\
Various macroscopic quantities of interest may be expressed as moments of $\F$ in the form $\int \psi(\vv) \F(\vv)\, d\vv$. 
For example, $\psi(\vv)=1$ yields the mass density $\rho(\vx,t) = \int \F(\vv; \vx,t)\, d\vv$, setting $\psi(\vv) = v_j$ yields the bulk velocity $u_j(\vx,t) = \int v_j \F(\vv; \vx,t)\, d\vv/\rho(\vx,t)$, and $\psi(\vv) = \frac{1}{2}\sum_k (v_k-u_k)^2$ yields the kinetic energy $e(\vx,t)=\frac{1}{2}\int \sum_k (v_k-u_k)^2 \F(\vv; \vx,t)\, d\vv$. Furthermore, the kinetic temperature $T$ is related to the kinetic energy via Boltzmann constant $k_b$, i.e. $T=\int \sum_k (v_k-u_k)^2 \F(\vv; \vx,t)\, d\vv/(3n k_b) $ where $n=\rho/m$ denotes the number density.
To simplify notation, we denote the fluctuating velocity by $\vw = \vv-\vu$, which implicitly depends on $\vx$ and $t$.
The Boltzmann operator satisfies conservation of mass, momentum, and energy, that is for $\mathcal{S}=\mathcal{S}^\text{Boltz}$,
\begin{align}
	\int \psi(\vv) \, \mathcal{S} \F(\vv; \vx,t)\, d\vv &= 0, 
	\qquad \psi(\vv)\in \left\{1, v_j, \tfrac{1}{2}\sum_k w_k w_k \right\}.\label{eqn:conservation}
\end{align}
Further, higher-order moments can be useful in describing the density $\F$. 
Of particular physical relevance are the pressure tensor $\bm p$ and heat flux $\bm q$, given by
\begin{align}
	p_{ij} = \int w_{i}w_{j} \F(v)\, dv,
	\qquad q_i = \frac{1}{2}\int w_i \sum_k w_k w_k \F(\vv)\, d\vv. \label{eqn:Boltz-conservation}
\end{align}
The deviatoric part of the pressure tensor $p_{\langle ij\rangle}$ with a negative sign gives us the stress tensor.
These quantities are not conserved by the collision operator, but rather relax towards their equilibrium values.
For Maxwell molecules, the relaxation rates of stress tensor and heat flux are found to be \citep[5.3.3]{Struchtrup2005}
\begin{align}
	\begin{split}
	\int w_{\langle i} w_{j\rangle} \mathcal{S}^\text{Boltz} \F(\vv)\, d\vv     &= -\alpha\, \rho\, p_{\langle ij\rangle} \\
	\tfrac{1}{2}\int w_i \sum_k w_k w_k \mathcal{S}^\text{Boltz} \F(\vv)\, d\vv &= - \frac{2}{3} \alpha\, \rho\, q_i,
	\end{split} \label{eqn:Boltz-relaxation}
\end{align}
for some $\alpha >0$.
In particular, the ratio of these relaxation rates is constant, yielding the Prandtl number $\tau_\text{Boltz} = {2}/{3}$.

\subsection{Fokker-Planck model}

To overcome the poor scaling of the collisions in DSMC, \cite{Jenny2010} suggested to decouple the flight paths of the particles and to replace the pairwise collisions by independent stochastic movement.
In their model, the state $(\vx(t), \vv(t))$ of a single particle evolves according to the Itô stochastic differential equation
\begin{align}
	\begin{split}
	d \vx(t) &= \vv(t) dt,\\
	d \vv(t) &= \vmu dt + \vF(\vx(t), t)dt + \mSigma^\frac{1}{2} d\vW(t),
	\end{split}\label{eqn:Ito}
\end{align}
where $\vW(t)$ is a standard Brownian motion, the covariance is isotropic and given by $\mSigma_{ij} = \frac{b}{3} e(\vx(t),t) \delta_{ij}$ for some $b\geq 0$, and the mean-reverting drift term $\mu_i= -a[v_i-u_i(\vx(t),t)]$.
The only interaction of the particles is via the mean-field quantities $\vu(\vx,t)$ and $e(\vx,t)
$, which are unknown in practice but may be approximated by a suitable averaging of the particle ensemble.
Letting the number of computational particles tend to infinity, the corresponding population density $\F$ satisfies the kinetic equation \eqref{eqn:kinetic}, with the right hand side
\begin{align*}
	\mathcal{S}\F(\vv; \vx,t) = \mathcal{S}^\text{FP}\F(\vv; \vx,t) &= \sum_i \frac{\partial}{\partial v_i} \left(-\mu_i \F \right) + \frac{1}{2}  \sum_{i,j} \frac{\partial^2}{\partial v_i \partial v_j}\left(\Sigma_{ij} \F\right) .
\end{align*}
The operator $\mathcal{S}^\text{FP}$ satisfies conservation of mass, momentum and energy as in \eqref{eqn:conservation} upon specifying $b=4a$.
While the Fokker-Planck operator has been suggested as an approximation of $\mathcal{S}^\text{Boltz}$ \cite{Jenny2010}, the outcome solution admits the wrong Prandtl number. Hence, 
for a collision operator $\mathcal{S}\neq \mathcal{S}^\text{Boltz}$ to yield a satisfactory approximate model for the Boltzmann equation, it should closely match the evolution of relevant higher order moments. 
That is, we would like to have $\int \psi(\vv) \mathcal{S}\F(\vv)\, d\vv = \int \psi(\vv) \mathcal{S}^\text{Boltz}\F(\vv)\, d\vv$ for $\psi \in \{ \psi_\alpha: \alpha=1,\ldots, M \}$ for some set of moments.
For the physically interesting cases of heat flux $q_i$ and stress tensor $p_{\langle ij\rangle}$, the linear Fokker-Planck model of \cite{Jenny2010} yields (recall $w_i=v_i-u_i$)
\begin{align}
	\begin{split}
	\int w_{\langle i} w_{j\rangle} \mathcal{S}^\text{FP} \F(\vv)\, d\vv     &= -2a\, p_{\langle ij\rangle} \\
	\frac{1}{2} \int w_i \sum_k w_k w_k \mathcal{S}^\text{FP} \F(\vv)\, d\vv &= - 3a\, q_i.
	\end{split} \label{eqn:FP-relaxation}
\end{align}
The relaxation rates may be adjusted by specifying the value of $a$.
For example, the rate for the stress tensor matches the evolution \eqref{eqn:Boltz-relaxation} of the Boltzmann model upon setting $a = {\alpha\,\rho}/{2}$, thus introducing an additional mean-field interaction via the mass density $\rho(\vx, t)$.
Just as for the Boltzmann operator, the ratio of the relaxation rates in \eqref{eqn:FP-relaxation} is constant and yields the Prandtl number $\tau_\text{FP}={3}/{2}\neq \tau_\text{Boltz}$.
Hence, the linear Fokker-Planck model may not match the evolution of both, stress tensor and heat flux, simultaneously.
\\ \ \\
To fix the issue with the wrong Prandtl number of the linear Fokker-Planck model, \cite{Gorji2011} changed the drift term $\vmu$ to include a cubic nonlinearity, see also \cite{Gorji2014} and \cite{Gorji2015}.
In principle, by fine-tuning the drift term, this approach could be extended to yield correct relaxation rates for higher-order moments.
A shortcoming of the original cubic Fokker-Planck model is that evaluation of nonlinear drift coefficients depends on estimation of moments up to fifth order which can introduce further error in noisy settings. Furthermore, it does not necessarily satisfy the H-theorem, i.e.\ for the corresponding cubic Fokker-Planck operator $\mathcal{S}^\text{CFP}$, it might occur that $\int \log(\F(\vv)) \mathcal{S}^\text{CFP} \F(\vv)\, d\vv <0$.
Recently, \cite{Gorji2019} showed that entropy can in fact be ensured to be increasing if the nonlinearity of the drift term $\vmu$ and the corresponding isotropic diffusion matrix $\mSigma$ are chosen carefully. 
\\ \ \\
A different approach to fix the issue with the Prandtl number is presented by \cite{Mathiaud2016}, who suggest to maintain the linear drift term and use a non-isotropic diffusion matrix $\mSigma$.
They show that the choice $\Sigma_{ij} = ({5}/{2}) a p_{ij} - ({3}/{2}) a \sum_k p_{kk} \delta_{ij}$ yields the correct Prandtl number $\tau=\tau_\text{Boltz}={2}/{3}$, and the H-theorem is satisfied.
The advantage of this approach compared to the approach of \cite{Gorji2019} is that the stochastic differential equation \eqref{eqn:Ito} admits an analytic solution because the drift is linear.
However, the approach of \cite{Mathiaud2016} suffers from the fact that the specified diffusion matrix $\mSigma$ might lack positive-definiteness.
If this is the case, a different diffusion matrix needs to be employed, leading to a wrong Prandtl number.
The cubic Fokker-Planck model and the ellipsoidal model suggested by \cite{Mathiaud2016} have been compared empirically by \cite{Jun2019}.

\section{Jump-diffusion particle methods}
\label{sec:jump_diffusion}

As our main result, we demonstrate that the evolution of higher-order moments of the Fokker-Planck particle method may also be corrected by introducing jumps to the velocity path $\vv(t)$.
In particular, our jump process will be simpler than the velocity jumps due to collisions in the exact Boltzmann equation.
To this end, we extend model \eqref{eqn:Ito} and let the state of a particle evolve according to the jump-diffusion model
\begin{align}
	\begin{split}
	d \vx(t) &= \vv(t) dt,\\
	d \vv(t) &= \vmu dt + \vF(\vx(t), t)dt + \mSigma^\frac{1}{2} d\vW(t) + \int \vc(\vz, \vx(t),t)\, N(d\vz, dt),
	\end{split}\label{eqn:JD}
\end{align}
where $N(d\vz,dt)$ is a Poisson random measure with intensity measure $\nu(d\vz)\,dt$.
The variable $\vz$ is called the mark (see the supplement), and $\vz\mapsto \vc(\vz,\vx,t)$ is the transfer function, mapping a mark $\vz$ to the jump size $\vc(\vz,\vx,t)$. 
Given any set $A\subset \R^3$, the measure $\nu(A)$ describes the expected number of jumps with mark $\vz\in A$ per unit of time.
Instead of working with the transfer function explicitly, it might be more intuitive to consider the local intensity measure $\nu(d\vc, \vx,t)$, which is defined as
\begin{align*}
    \nu(A, \vx,t) = \nu\left( \{ \vz: c(\vz,x,t)\in A \} \right).
\end{align*}
Then $\nu(A, \vx,t)$ describes the expected number of jumps of size $\vc\in A$ per unit of time, for a particle located at $\vx$ at time $t$.
For the model \eqref{eqn:JD} to be sensible, we require that $\int \min(1,\|\vc(\vz, \vx,t)\|)\nu(d\vz) = \int \min(1, \|\vc\|) \nu(d\vc,\vx,t) <\infty$.
A detailed introduction to jump-diffusion models of the form \eqref{eqn:JD} is given in the appendix of this article.
\\ \ \\
In order to use model \eqref{eqn:JD} as a particle scheme to approximate Boltzmann's equation, we need to study the evolution of the corresponding particle density.
As outlined in the appendix, it satisfies equation \eqref{eqn:kinetic} with collision operator
\begin{align*}
	\mathcal{S} \F (\vv) = \mathcal{S}^\text{JD} \F(\vv) &= \mathcal{S}^\text{FP} \F(\vv) + \mathcal{S}^\text{J} \F(\vv),\\
	\mathcal{S}^\text{J} \F(\vv; \vx, t) &= \int \left[\F(\vv-\vc(\vz, \vx,t)) - \F(\vv)\right]\, \nu(d\vz) \\
	&= \int \left[\F(\vv-\vc) - \F(\vv)\right]\, \nu(d\vc, \vx,t).
\end{align*}
The latter integral is in particular finite if $\vv\mapsto \F(\vv)$ is Lipschitz continuous and bounded, and $\int \min(1,\|\vc\|)\nu(d\vc,\vx,t)<\infty$, as assumed.
This evolution equation should be interpreted only formally.
Additional regularity requirements are necessary to make the evolution of $\F$ mathematically precise, which is however out of scope of this article.
\\ \ \\
The evolution of moments $\psi(\vv)$ may be determined as
\begin{align}
	\int \psi(\vv)\mathcal{S}^\text{J} \F(\vv; \vx, t)\, d\vv 
	&= \int \int \left[\psi(\vv + \vc) - \psi(\vv)\right] \nu(d\vc, \vx, t) \F(\vv; \vx,t)\, d\vv. \label{eqn:moment-jumps}
\end{align} 
Since the local intensity measure $\nu(d\vc, \vx, t)$ is an infinite dimensional object, the detailed specification of \eqref{eqn:JD} admits sufficiently many degrees of freedom to closely match the Boltzmann collision operator.

\subsection{Boltzmann equation as a jump-diffusion}
\label{sec:Bolt_as_jump}

In fact, we may even specify $\nu$ such that the moment evolution of the Boltzmann operator is matched exactly.
It holds that \citep[eq.\ 3.28]{Struchtrup2005}
\begin{align}
	&\quad \int \psi(\vv)\mathcal{S}^\text{Boltz} \F(\vv)\, d\vv \nonumber \\
	&= \int \left[\psi(\vv') - \psi(\vv)\right] \F(\bar{\vv})\F(\vv) |\vv - \bar{\vv}| I(\Omega, |\vv - \bar{\vv}|) \, d\Omega\, d\bar{\vv} \, d\vv \nonumber \\
	&= \int \left\{ \int \left[\psi(\vv + \vc(\vv, \bar{\vv},\Omega)) - \psi(\vv)\right] \F(\bar{\vv}) |\vv-\bar{\vv}| \,I(\Omega, |\vv-\bar{\vv}|)\, d\Omega\,d\bar{\vv} \right\} \F(\vv)\, d\vv \nonumber \\
	&= \int \left\{ \int \left[\psi(\vv + \vc) - \psi(\vv)\right] \nu^\text{Boltz}(d\vc; \vv, \vx, t) \right\} \F(\vv; \vx, t)\, d\vv. \label{eqn:boltzmann-jump-inversion}
\end{align}
Here, $\vc(\vv, \bar{\vv}, \Omega) = \vv'-\vv$ is the change of velocity due to collision with a particle with velocity $\bar{\vv}$ and collision angle $\Omega$.
The measure $\nu^\text{Boltz}$ on $\R^3$ is given by
\begin{align*}
    \nu^\text{Boltz}(A;\vv,\vx,t) = \int \mathds{1}\left\{\vc(\vv,\bar{\vv},\Omega)\in A\right\}\; \F(\bar{\vv})\cdot |\vv-\bar{\vv}| \cdot I(\Omega, |\vv-\bar{\vv}|)\, d\Omega\, d\bar{\vv}.
\end{align*}
Since the identity \eqref{eqn:boltzmann-jump-inversion} holds for arbitrary moment functions $\psi(\vv)$, we conclude that
\begin{align*}
	\mathcal{S}^\text{Boltz} \F(\vv; \vx, t) &= \int \left[\F(\vv-\vc) - \F(\vv)\right]\nu^\text{Boltz}(d\vc; \vv, \vx,t) = \mathcal{S}^\text{J}\F(\vv; \vx,t)~.
\end{align*}
This match with the Boltzmann operator suggests to build a particle Monte Carlo scheme by simulating particles according to \eqref{eqn:JD} with $\vmu=0$, $\mSigma=0$, and jump measure $\nu^\text{Boltz}$.
Since $|\nu^\text{Boltz}| = \nu^\text{Boltz}(\R^3)$ is finite, the process \eqref{eqn:JD} has finitely many jumps and may be sampled numerically by a suitable Euler scheme.  
The value $|\nu^\text{Boltz}|$ is the expected total number of jumps per time unit, which directly corresponds to the number of collisions in Boltzmann's equation.
Hence, in dense regimes, the process \eqref{eqn:JD} with jump measure $\nu^\text{Boltz}$ incurs many jumps.
Since the form of the Boltzmann jump measure $\nu^\text{Boltz}$ is rather generic, we may not expect to find a fast sampling procedure for the corresponding jump process.
Instead, all jumps need to be resolved individually, and hence this Boltzmann jump-diffusion model suffers from computational limitations similar to the DSMC method.

\subsection{The Gamma-Boltzmann model}
\label{sec:Gamma-Boltzmann}

Fortunately, there exist jump measures $\nu$ which allow for more efficient sampling of the process \eqref{eqn:JD}, at the price of matching only finitely many moments of the Boltzmann operator.
We suggest to use an intensity measure corresponding to the Gamma process, which is given by
\begin{align}
	\nu^\Gamma(d\vc,\vx,t) & = \sum_i \gamma_i\frac{\exp(-c_i/\lambda_i)}{|c_i|} {\mathds{1}}(\lambda_i c_i >0)\, S_{\R_i}(d\vc), \label{eqn:nugamma}
\end{align} 
where $S_{\R_i}(d\vc)$ denotes the Lebesgue measure on the $i$-th axis, i.e.\ $S_{\R_1}(d\vc)$ is the one-dimensional Lebesgue measure on $\R_1=\R\times\{0\}\times \{0\}$.
The parameters $\gamma_i$ and $\lambda_i$, which may depend on $\vx$ and $t$, satisfy $\gamma_i\geq 0$ and $\lambda_i\in\R\setminus\{0\}$.
This local intensity measure may be realized by choosing $\nu(d\vz)$ and the transfer function $\vc^\Gamma(\vz, \vx, t)$ suitably.
\\ \ \\
We highlight that the intensity measure $\nu^\Gamma$ is infinite, which implies that the corresponding velocity trajectory \eqref{eqn:JD} has infinitely many jumps.
Since $\nu^\Gamma$ concentrates around the origin, the majority of these infinitely many jumps are very small, such that the trajectory is still well-defined.
In fact, since $\int \min(1, \|\vc\|)\, \nu^\Gamma(d\vc,\vx,t)<\infty$, the jumps are actually summable, see the appendix.
Moreover, the measure $\nu^\Gamma$ is only supported on the axis, which implies that each individual jump only affects a single dimension.
\\ \ \\
We suggest to instantiate model \eqref{eqn:JD} by choosing values $b,c\geq 0$ such that $b+c=2a$, and setting
\begin{flalign}
    \begin{split}
	\mu_i       &= -a (v_i-u_i) - \lambda_i \gamma_i,\\
	\Sigma_{ij} &=  \delta_{ij} \frac{\sum_{k} p_{kk}}{3} \frac{1}{\rho} \left[ b + c \mathds{1}_{q_i=0} \right] \\
	\lambda_i   &= \frac{5a}{c} \frac{q_i}{\sum_k p_{kk}}, \\
    \gamma_i    &= \frac{c^3}{75 a^2} \frac{1}{\rho} \frac{(\sum_k p_{kk})^3}{q_i^2} \mathds{1}_{q_i\neq 0}.\label{eqn:JD-full}
    \end{split}
\end{flalign}
With this specification, the jump-diffusion operator $\mathcal{S}^\text{JD}$ conserves mass, momentum, and energy
\begin{align*}
    \int 1\cdot \mathcal{S}^\text{JD}\F(\vv; \vx,t)\, d\vv =0, \quad
    \int v_i \cdot \mathcal{S}^\text{JD}\F(\vv; \vx,t)\, d\vv =0, \quad
     \int \sum_k w_k w_k \mathcal{S}^\text{JD} \F(\vv)\, d\vv = 0.
\end{align*}
Furthermore, as derived in equation (A.4) in the appendix, the evolution of the stress tensor $p_{\langle i j \rangle}$ and the heat flux $q_i$ are given by
\begin{align}
    \begin{split}
	\int w_{\langle i} w_{j\rangle} \mathcal{S}^\text{JD} \F(\vv)\, d\vv 
	&= -2a p_{\langle ij\rangle} ,\\
	\frac{1}{2}\int w_i \sum_k w_k w_k \mathcal{S}^\text{JD}\F(\vv)\, d\vv 
	&= -\frac{4}{3}a q_i. \label{eqn:JD-relaxation}
	\end{split}
\end{align}
Hence, the model gives rise to the correct Prandtl number $\tau_\text{JD} = \tau_{\text{Boltz}} = {2}/{3}$, for any choice of $c\in(0,2a]$.
\\ \ \\
The Fokker-Planck model of \cite{Jenny2010} corresponds to the special case $c=0$.
Hence, the introduction of the jump component, $c>0$, is crucial to ensure the correct Prandtl number.
Compared to purely Gaussian noise, the jumps have a bigger impact on the higher order moments of the particle velocities.
In particular, the relaxation of the heat flux $q_i$, as a third-order moment, is diminished due to the jumps.
To achieve this, the precise value of $c>0$ is not important because the effect on the third order moments may be achieved by various combinations of $\lambda_i$ and $\gamma_i$. 
That is, our specification as a function of $c$ leads to the correct Prandtl number for any choice $c\neq0$.
This holds true for constant values $c$, but also if $c=c(\vx,t)$ is a function of the solution $\F(\vv; \vx,t)$ itself. 
Also, for any choice of $c$, we find that $\mathcal{S}^\text{JD}\F=0$ if and only if $\F$ is the Maxwellian equilibrium distribution; see Section A.3 in the appendix.

\section{Particle Monte Carlo scheme}\label{sec:sampling}
Here, similar to DSMC and cubic FP, we consider samples of distribution function and evolve their positions and velocities in two separated steps of streaming and velocity update. Holding the moments constant during a time step, the evolution of the particle velocity in the Gamma-Boltzmann model may be simulated exactly, as demonstrated below. We also present an approximation which may be helpful in regimes where the exact solution becomes computationally demanding. An efficient solution algorithm combining the exact and approximated solution to the Gamma-Boltzmann model is provided in  \S~\ref{sec:sol_alg}.

\subsection{Exact solution of particle velocity}\label{sec:gamma-sampling}

If we keep the local moments constant for the interval $[t,t+\Delta]$, then the jump measure $\nu(d\vz;v,x,t)=\nu^\Gamma(d\vz)$ is constant as well. 
In particular, $\int_t^{t+r} \vc(\vz, X(t),t) N(d\vz, ds) = \mathbf{J}(r)$ 
is a Lévy process, namely a Gamma process, see above. 
Hence, the velocity of a single particle evolve according to the jump-diffusion
\begin{align*}
    dv_i(t+r) &= \left[-a(v_i(t+r) - u_i(t)) - \lambda_i \gamma_i\right] dt + [b+c\mathds{1}_{q_i=0}] \frac{\sum_k p_{kk}}{3} dW_i(t+r) + dJ_i(r),
\end{align*}
which admits the analytical solution
\begin{align}
    \begin{split}
    v_i(t+r) 
    &= (1-e^{-a r}) u_i(t) + e^{-a r}v_i(t) - \lambda_i \gamma_i r \\
    &\quad + [b+c\mathds{1}_{q_i=0}] \frac{\sum_k p_{kk}}{3}\int_0^r e^{-a(r-s)} \, dW_i(t+s) + \int_0^r e^{-a(r-s)} \, dJ_i(s).
    \end{split}
    \label{eq:ex_eq1}
\end{align}
The first three terms are deterministic and fully explicit. 
The second term has a multivariate normal distribution with covariance $\Sigma_{ij} = \delta_{i j} [b+c\mathds{1}_{q_i=0}] \frac{\sum_k p_{kk}}{3} \frac{1-e^{-2ar}}{2a}$. 
\\ \ \\
The last term is a stochastic integral w.r.t.\ a Lévy process and does not admit a simple closed form solution.
Nevertheless, we may utilize the exact simulation scheme of \cite[Algorithm 4.1]{Qu2019} to find that\footnote{The algorithm of \cite{Qu2019} contains an error and is only correct for $\rho=1$. This is not a restriction because in the model formulation of Qu et al., the parameter $\beta$ and $\rho$ serve the same purpose, i.e.\ their model is overparametrized.}
\begin{align}
    \int_0^r e^{-a(r-s)} \, d J_i(s) 
    \overset{d}{=} 
    \Gamma_r + \sum_{k=1}^{N_r} S_{k,r}, \label{eqn:OU-gamma}
\end{align}
where the $\Gamma_r, N_r, S_{k,r}$ are independent random variables such that
\begin{align*}
    \Gamma_r &\sim \Gamma(\gamma_i r, \tfrac{1}{\lambda_i} e^{a r}), \\
    N_r &\sim \mathrm{Poi} (\tfrac{1}{2}\gamma_i a r^2), \\
    S_{k,r} & \sim \text{Exp}(\tfrac{1}{\lambda_i} e^{a r \sqrt{U}}), \quad U \sim U(0,1),
\end{align*}
i.e.\ the $S_{k,r}$ are mixed exponentially distributed.
This formula is valid for $\lambda_i>0$, otherwise consider $-J(r)_i$.
\\ \ \\
Note that the expectation of $N_r$ is $\mathbb{E}(N_r) =\tfrac{1}{2}\gamma_i a r^2$, hence the computational cost to evaluate \eqref{eqn:OU-gamma} is on average $\mathcal{O}(1+\gamma_i r^2)$.
We will usually set $r=\Delta$.
However, if $\gamma_i$ is very large, we might want to choose $r< \Delta$ and perform multiple exact steps using \eqref{eqn:OU-gamma}.
If we split the interval $[t,t+\Delta]$ in $m$ sub-intervals of equal length, the computational effort will be on average $\mathcal{O}(m(1+\gamma_i \Delta^2/m^2)) = \mathcal{O}(m + \frac{\gamma_i \Delta^2}{m})$.
Thus, the optimal choice of $m$ will be $m\approx \lceil\Delta \sqrt{\gamma_i}\rceil$. 
If $\Delta$ is sufficiently small, $m=1$ will usually be satisfactory, but in some extreme cases the described variant might be useful.

\subsection{Approximate solution to particle velocity}\label{sec:gamma-sampling-approx}

The representation \eqref{eqn:OU-gamma} may also be used to derive an approximate numerical scheme for the regime where $\gamma_i$ is large. We use that $S_{k,r} = \lambda_i e^{-ar \sqrt{U_{k,r}}} Z_{k,r}$, where $U_{k,r}$ are independent and identically distributed (iid) standard uniform random variables, and $Z_{k,r}$ are iid standard exponential random variables.
This suggests the approximation
\begin{align}
    \int_0^r e^{-a(r-s)} \, d J_i(s) 
    &\overset{d}{=} 
    \Gamma_r + \lambda_i \E(e^{-ar \sqrt{U}}) \sum_{k=1}^{N_r} Z_{k,r} + \epsilon_r, \label{eqn:OU-gamma-approx}
\end{align}
where $\epsilon_r$ denotes the approximation error. The advantage of this scheme is that the sum may be aggregated, since the sum of independent exponential random variables follows a Gamma distribution, i.e.
\begin{align*}
    \lambda_i \E(e^{-ar \sqrt{U}}) \sum_{k=1}^{N_r} Z_{k,r} 
    &\overset{d}{=} \Gamma\left(N_r, \frac{1}{\lambda_i \E(e^{-ar \sqrt{U}})}\right), \qquad N_r \sim \text{Poi}(\tfrac{1}{2} \gamma_i a r^2).
\end{align*}
Hence, the sum admits a mixed Gamma distribution, which may be simulated efficiently, even in the critical regime $\gamma_i$ is very large. We also remark that $\E(e^{-ar \sqrt{U}}) = 2\frac{e^{ar}}{(ar)^2} (e^{ar} -1 - ar)$.
\\ \ \\
In order to analyze the error $\epsilon_r$, note that $|\E(e^{-ar\sqrt{U}}) - e^{-ar\sqrt{U}}| \leq ar$. Since the summands are independent, we conclude that 
\begin{align*}
    \E(\epsilon_r)=0, \qquad \E ( |\epsilon_r|^2 ) = \Var(\epsilon_r) \leq \lambda_i^2 (ar)^2 \E(N_r) = \frac{a^3r^4}{2} \gamma_i\lambda_i^2.
\end{align*} 
By our model specification \eqref{eqn:JD-full}, the regime $\gamma_i\to\infty$ corresponds to low heat flux $q_i\to 0$. But in this regime, the product $\gamma_i\lambda_i^2$ stays bounded. Hence, the approximate scheme \eqref{eqn:OU-gamma-approx} yields a satisfactory approximation in situations where the exact scheme is prohibitively expensive.

\subsection{Solution algorithm}
\label{sec:sol_alg}
In this section, we provide a detailed solution algorithm, i.e. Algorithm \ref{alg:jump_process_algorithm}, that solves the Gamma-Boltzmann model for future reference. First, similar to other particle methods, one needs to discretize the phase space with $N_p$ particles, i.e.
\begin{flalign}
    \mathcal{F}(\bm v,\bm x, t) = \lim_{N_p\rightarrow \infty} \sum_{j=1}^{N_p} {w^{(j)}} \delta( \bm v^{(j)}(t)-\bm v ) \delta( \bm x^{(j)}(t)-\bm x )  
\end{flalign}
where $w^{(j)}$ is the weight associated with the $j$th particle,
and $\delta(.)$ is the Dirac delta function. Having discretized the solution domain in $\bm x$ dimension into $N_\mathrm{cells}$ cells, a constant weight for all particles leads to the trivial computation of density and number density for the $i$th cell, i.e. $\rho^{(i)} \approx w N_{p/\mathrm{cell}}$ and $n^{(i)}=\rho^{(i)}/m$. The fixed value of particle weight is initially set given the initial mass density of the system $\rho_0$, volume of the system, and the number of particles in the domain. Since in practice we can only deploy a finite number of samples, the stochastic representation is subject to statistical errors.
\\ \ \\
In order to avoid high computational cost associated with performing all the jumps exactly,  we estimate the cost associated with jumps and deploy the approximate solution, see \S~\ref{sec:gamma-sampling-approx},  as the cost exceeds a given threshold $\epsilon$. The simulation results of this work are obtained by deploying this algorithm.
\\ \ \\
\begin{algorithm}[H]
\SetAlgoLined
 Initialize particles in the phase space\;
 \While{$t<t_\mathrm{final}$}{
  \For{$i=1,...,N_\mathrm{cells}$}{
  Compute needed moments\;
   $\tau^{(i)} = 2\mu/(n^{(i)} k_b T^{(i)})$ and $a^{(i)}={1}/{\tau^{(i)}}$\;
   Select a value for $c^{(i)}\in(0,2a]$, e.g. $c^{(i)}=a^{(i)}$\; 
   $b^{(i)} = 2a^{(i)}-c^{(i)}$ \;
   $\lambda_j^{(i)} = 5 a q_j^{(i)}/( c^{(i)} \sum_k p_{kk}^{(i)})$ \;
   $\gamma_j^{(i)}  = (c^{(i)})^3  (\sum_k p_{kk}^{(i)})^3 / (75 (a^{(i)})^2) / (q_j^{(i)})^2$\;
   \For{$j=1,...,N_\mathrm{p/cell}^{(i)}$}{
    \eIf{$\gamma_j^{(i)} \Delta t^2 < \epsilon$}{
        Evolve velocity $\bm v$ of particle $j$ according to Eqs.~\eqref{eq:ex_eq1}-\eqref{eqn:OU-gamma}\;
    }
    {
    Evolve velocity $\bm v$ of particle $j$ using the approximate solution Eq.~\eqref{eqn:OU-gamma-approx}\;
    }
   }
  }
  Stream particles with the new velocity\;
  Apply boundary condition\;
  $t=t+\Delta t$\;
 }
 \caption{Particle Monte Carlo scheme for the Gamma-Boltzmann approximation to the Boltzmann equation}
 \label{alg:jump_process_algorithm}
\end{algorithm}

\section{Computational results}
\label{sec:comp_res}
In this section, an implementation of the devised Gamma-Boltzmann model is compared to the analytical solution as well as benchmarks in several test cases. In \S~\ref{sec:hom_toy}, we consider the relaxation of a bi-modal distribution to equilibrium in a spatially homogeneous setting. This setup serves as a toy problem where we show that the measurement of relaxation rates is in agreement with the analytical derivation. 
\\ \ \\
Then, we test the solution obtained from the Gamma-Boltzmann model against DSMC and cubic FP model in Couette flow \S~\ref{sec:couette} and lid-driven cavity \S~\ref{sec:lid-driven}. Here, we take Argon as the monatomic hard-sphere gas with mass $m=6.6335 \times 10^{-26}\ \mathrm{kg}$, and viscosity $\mu = 2.117\times 10^{-5}\ \mathrm{kg.m^{-1}.s^{-1}}$ at $T_0=273\ \mathrm{K}$. In the result section, we refer to Knudsen number 
\begin{flalign}
    \mathrm{Kn} = \frac{\lambda}{L}
\end{flalign}
where $L$ denotes the length scale of problem and $\lambda$ is the mean free path of hard-sphere molecules. Furthermore, we deployed $r=\Delta t$ and $\epsilon=0.1$ in Algorithm~\ref{alg:jump_process_algorithm}, everywhere unless mentioned otherwise.

\subsection{Homogeneous toy example}
\label{sec:hom_toy}
As a proof of concept, we study a simple example where we assume the particle distribution to be perfectly homogeneous in $\R^d$, without boundaries. That is, $\F(\vv;\vx,t)=\F(\vv;t)$. 
We treat this case by simulating $N_p=10^6$ particles representing the distribution function, and consider the evolution equation for velocity only. We choose $c=2a$ and $b=0$, such that the Gaussian component is omitted. The relaxation rate is fixed by setting $a={1}/{2}$. At time $t=0$, we initialize the distribution as a mixture of two highly concentrated Gaussian distributions,
\begin{align*}
    \F(\vv;0) &= \frac{2}{3} \varphi_{0,\mathbf{B}}(\vv) + \frac{1}{3} \varphi_{\tau,\mathbf{B}}(\vv), \\
    B_{ij} &= 10^{-4} \delta_{ij}, \\
    \tau &= (3,6,9)
\end{align*}
Here, $\varphi_{\tau,\mathbf{B}}$ denotes the density of a multivariate normal distribution with mean value $\tau\in\R^3$ and covariance matrix $\mathbf{B}\in\R^{3\times 3}$. In particular, the initial velocity distribution is far from the Maxwellian equilibrium.
\\ \ \\
We simulate the particles in the interval $[0,10]$ and update the ensemble moments at step size $\Delta=10^{-2}$.
Since the exact scheme becomes computationally expensive for small value $\gamma_i$, we change the simulation method if $\gamma_i \Delta^2$ raises above a threshold of $0.1$. In this regime, we use the approximate scheme \ref{sec:gamma-sampling-approx} with smaller step size $r=10^{-3}$.

\begin{figure}
  \centering
\includegraphics[width=0.45\textwidth]{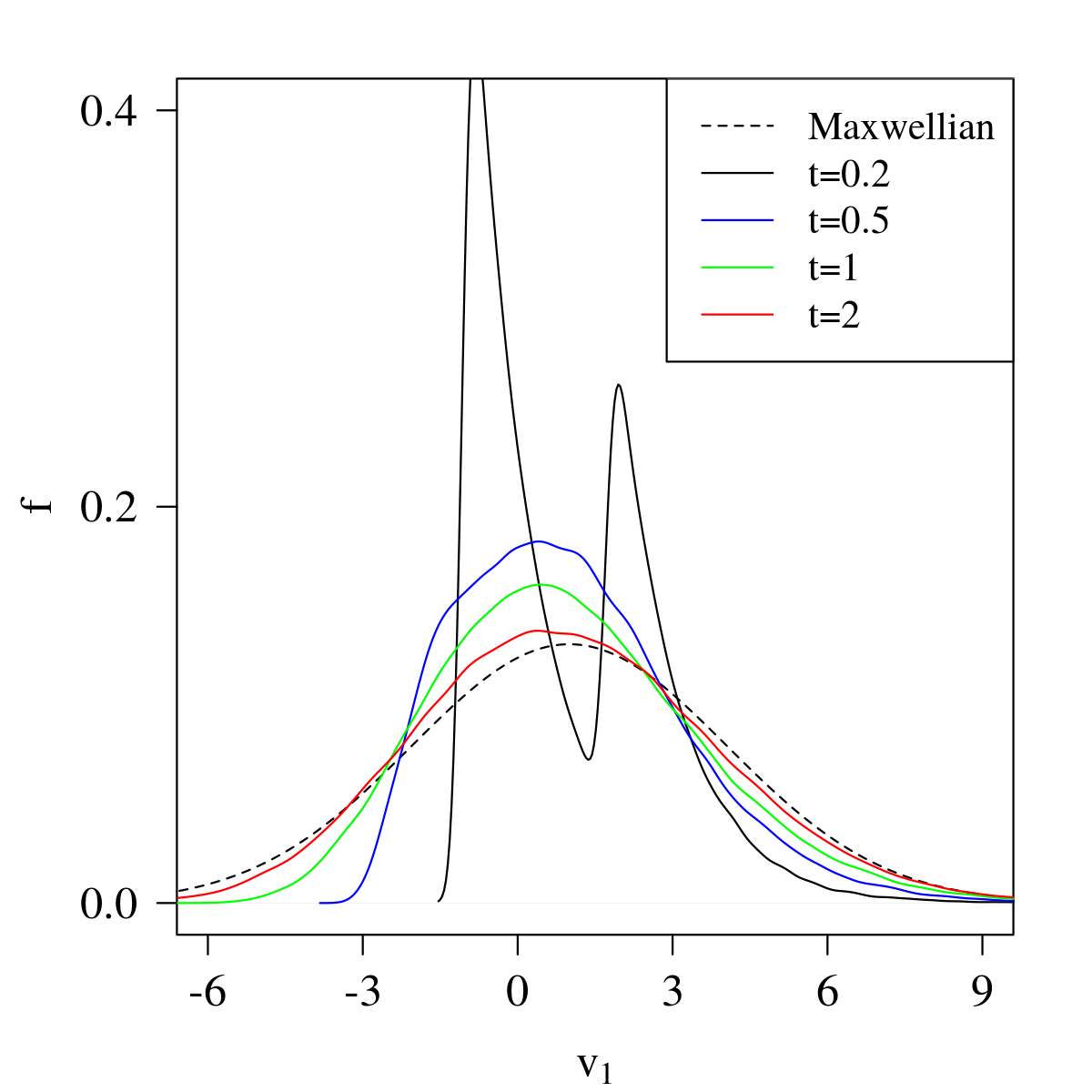}
  \caption{Evolution of a bi-modal distribution in $v_1$ dimension towards the Maxwellian distribution. The distribution is  estimated at $t=0.2,0.5,1,2$ by solving the Gamma-Boltzmann particle scheme.}
  \label{fig:homogen_conv}
\end{figure}

\begin{figure}
  \centering
\includegraphics[width=0.45\textwidth]{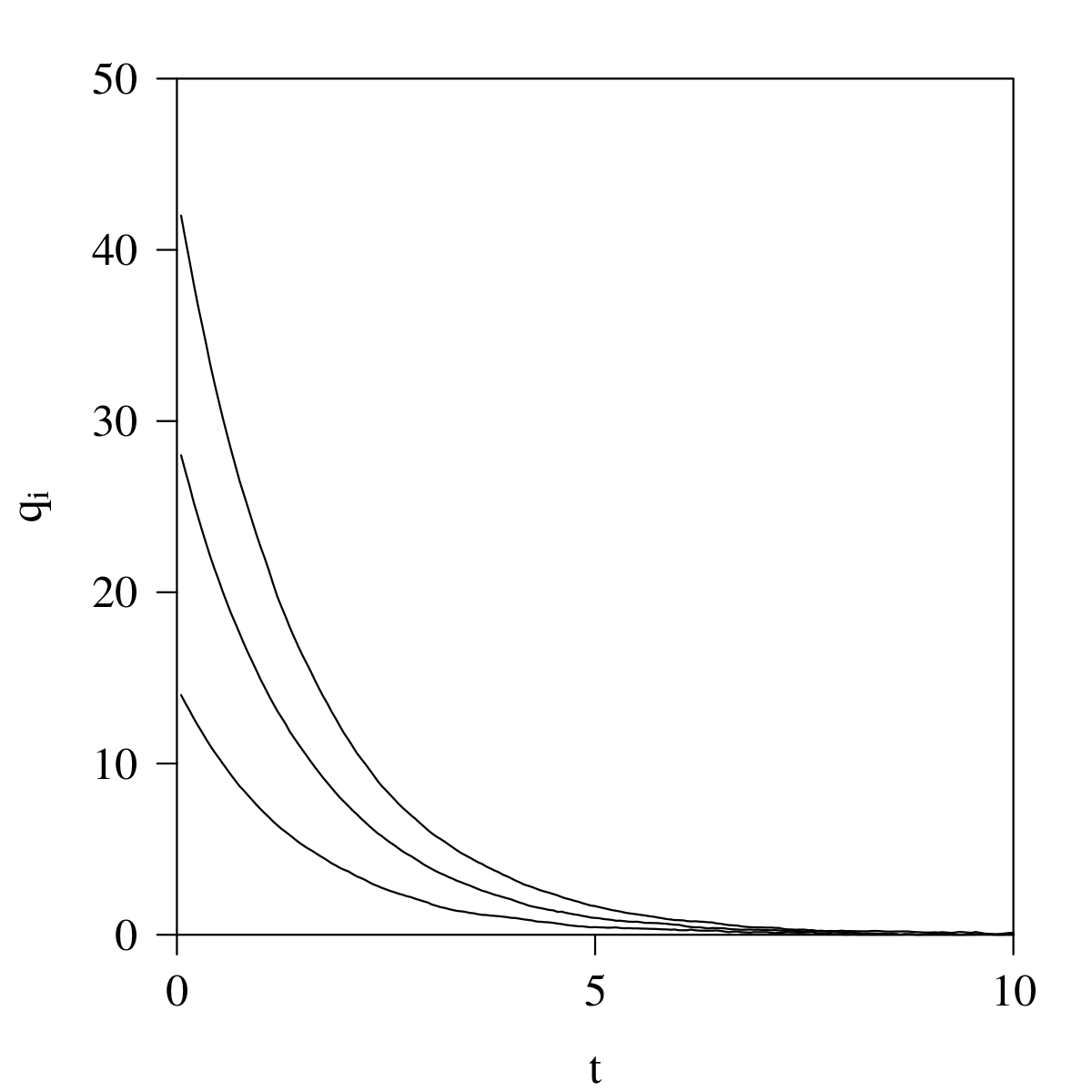}
\includegraphics[width=0.45\textwidth]{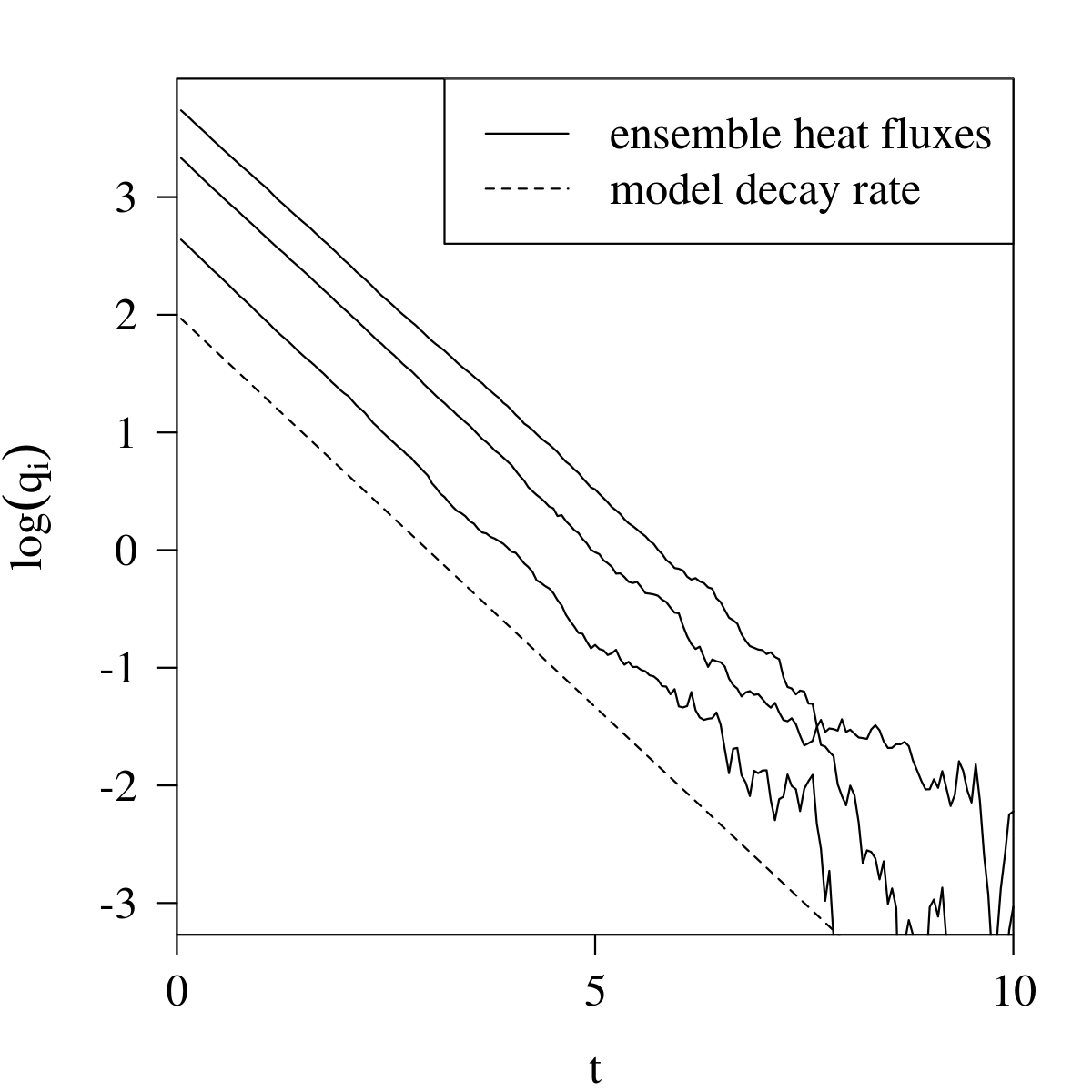}
  \caption{Relaxation of heat flux as the particle distribution approaches equilibrium. The heat flux is computed via the Gamma-Boltzmann particle scheme.}
  \label{fig:homogen_heatflux}
\end{figure}
\ \\
The evolution of the velocity distribution is depicted in Figure \ref{fig:homogen_conv}, in a single dimension. The convergence towards the Maxwellian equilibrium distribution is evident. In Figure \ref{fig:homogen_heatflux}, we depict the heat fluxes $q_1 < q_2 < q_3$ as a function of time. The exponential decay matches the theoretical model \eqref{eqn:JD-relaxation}. The logarithmic plot reveals that the decay rate $\frac{4}{3}a$ is attained in the beginning. The erratic behavior for large times may be explained by the sampling error incurred by approximating the equation with finitely many particles.

\FloatBarrier
\subsection{Couette flow}
\label{sec:couette}
\begin{figure}
  \centering
\includegraphics[scale=0.9]{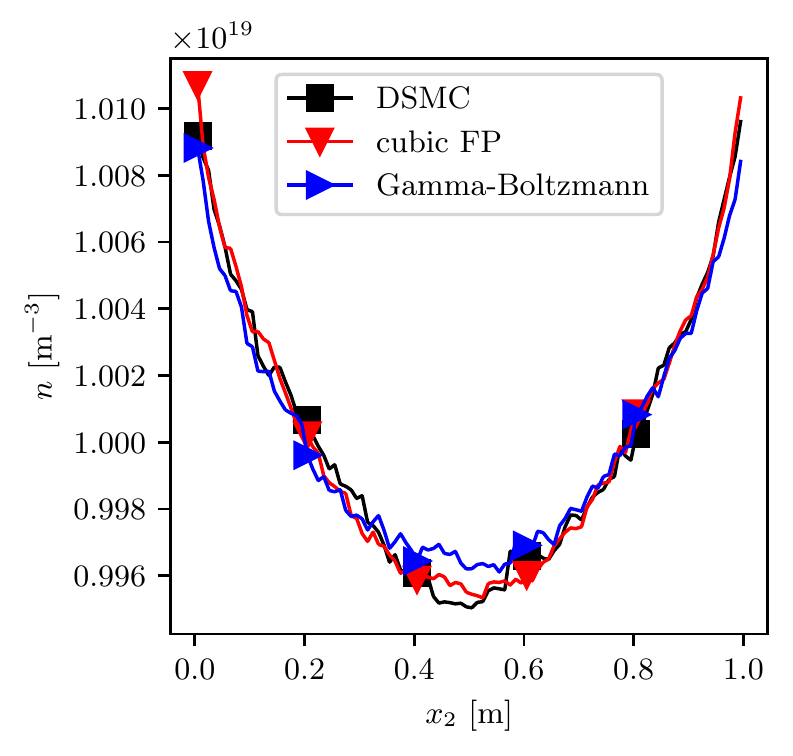}
\includegraphics[scale=0.9]{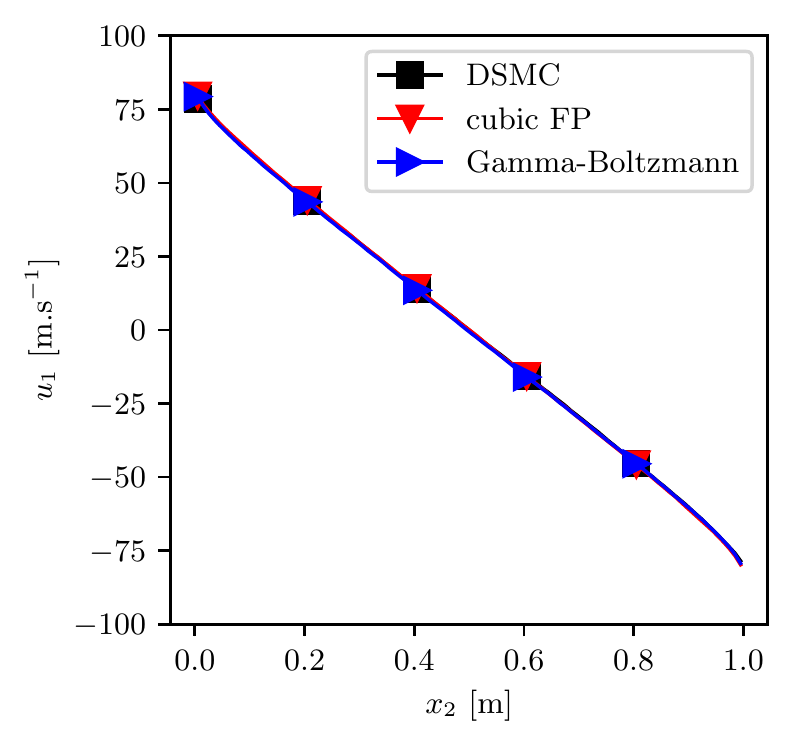}
\includegraphics[scale=0.9]{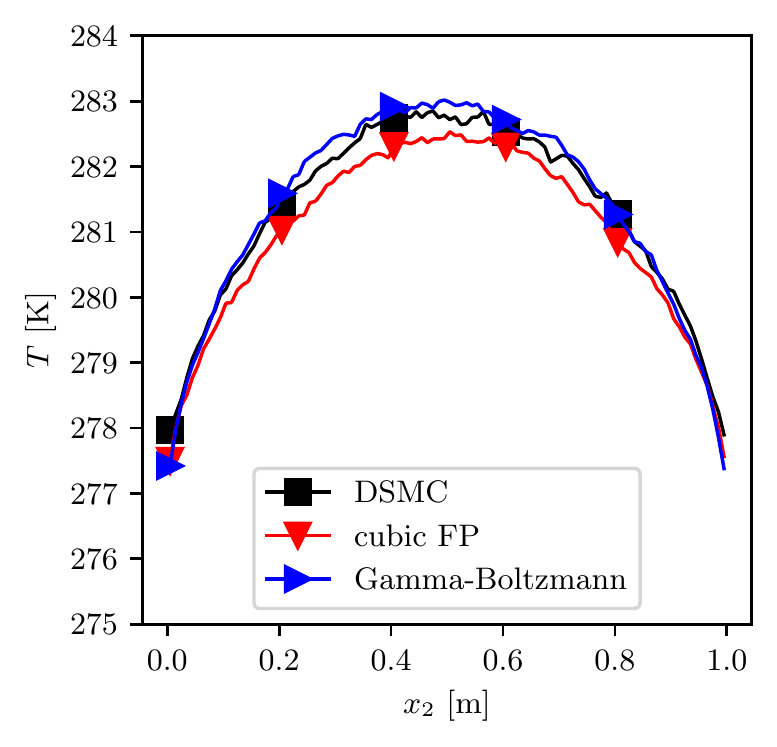}
\includegraphics[scale=0.9]{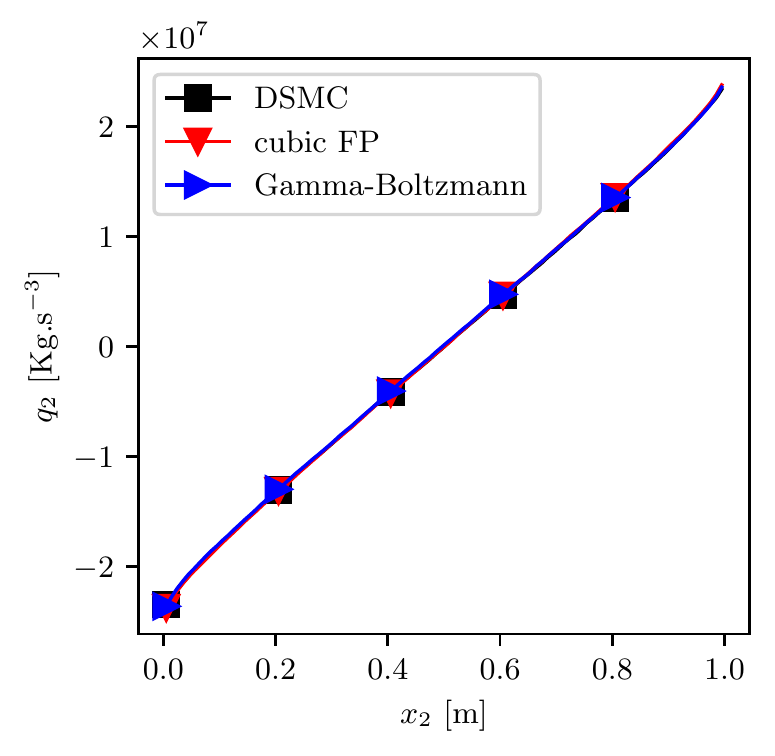}
  \caption{Profiles of number density, bulk velocity, temperature, and kinetic heat flux for  Couette flow between moving thermal walls at $100\ \mathrm{m.s^{-1}}$ in opposite directions obtained from DSMC, cubic FP model, and the Gamma-Boltzmann model, shown in black, red, and blue respectively. Here, the Knudsen number is $\mathrm{Kn}=0.17$.}
  \label{fig:couette}
\end{figure}

\begin{figure}
  \centering
\includegraphics[scale=1.0]{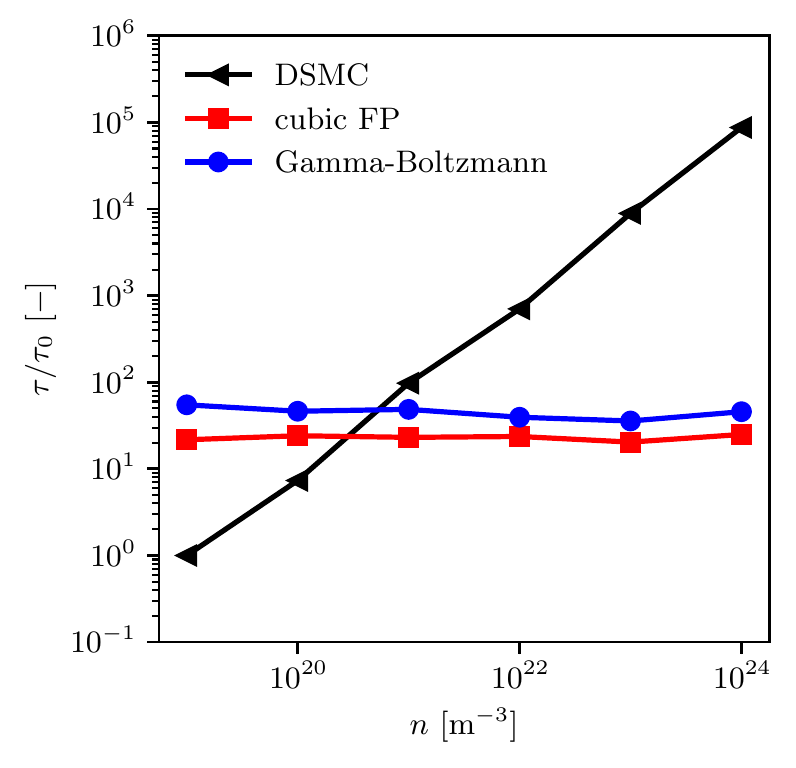}
  \caption{Normalized computation time $\tau$ of solving the Couette flow problem using DSMC, cubic FP model, and the Gamma-Boltzmann model against the initial number density of the gas using $N_{p/\mathrm{cell}}=1000$ particles per cell in all cases. The computation times are normalized with $\tau_0$, i.e., the one obtained from DSMC  for $n_0=10^{-19}\ \mathrm{m}^{-3}$. In all the simulations, time step size is $\Delta t = 10^{-6}\ \mathrm{s}$, and initial temperature is  $T_0=273\ \mathrm{K}$.}
  \label{fig:couette_cost}
\end{figure}

In order to investigate the accuracy of the devised model in a shear dominant setting, we simulate a planar Couette flow. Consider a particle system enclosed between two thermal moving walls with velocity $\bm u_\mathrm{w}=(\pm 100,0,0)^T\ \mathrm{m.s^{-1}}$ and temperature $T_\mathrm{w}=273\ \mathrm{K}$ at the distance of $L$ from one another where $x_2$ is normal to the walls. Hence, the solution domain is $x_2\in [0,L]$, while ignoring the other dimensions in $\bm x$, and the initial number density  $n_0=10^{19}\ \mathrm{m}^{-3}$ and initial temperature $T_0=273\ \mathrm{K}$. As particles hit the walls (leave the domain), we sample the velocity of the incoming particle from the flux of shifted Gaussian distribution and stream the particle with the new velocity for the remainder of the time step. For example, particles that enter the domain from the lower wall at $x_2=0$, the new velocity component normal to the wall is sampled from the flux of the Maxwellian distribution, i.e.\ the probability density of the sampled flux is proportional to
$v_2\mathcal{N}(0,k_b T_w/m),\ v_2>0$.
This distribution may be sampled explicitly as
\begin{flalign}
    v_2 = \sqrt{2 k_b T/m }  \sqrt{-\log(\alpha) }
\end{flalign}
where $\alpha\sim \mathcal{U}(0,1)$ is a uniformly distributed random variable.
In other directions, we sample $v_j \sim  \mathcal{N}(u_{w,j},k_b T_w/m)$ for $j=1,3$. Here, $\mathcal{N}( m, \sigma^2)$ is the normal distribution function with mean $ m$ and variance $ \sigma^2$, and the corresponding probability density $\varphi_{m,\sigma^2}$. Initially, particles are distributed uniformly in $x_2 \sim \mathcal{U}([0,L])$ and normally distributed in velocity $\bm v \sim \mathcal{N}(\bm{u}_0,k_b T_0/m \bm I)$, where $k_b$ is the Boltzmann constant. As particles evolve and hit the boundaries, the evolution of moments evolve and reach a steady state profile, i.e. a stationary solution for the distribution function in the solution domain is achieved. A convergence study lead us to use initially $N_\mathrm{p/cell}=1000$ particles per cell, the time step size of $\Delta t=10^{-6}$, and  $N_\mathrm{cells}=100$ computational cells in $x_2$.
\\ \ \\
Here, we simulate the Couette flow using Direct Simulation Monte Carlo (DSMC) and cubic Fokker-Planck model (FP) as benchmarks against the Gamma-Boltzmann model developed in this work. We deploy Algorithm~\ref{alg:jump_process_algorithm} in order to numerically solve the Gamma-Boltzmann model with $c=a$ where  $a ={1}/{\tau}$,
  $\tau = {2 \mu}/{p}$ is the time scale of diffusion part of the process and
$p=nk_bT$ is the equilibrium pressure of ideal gas. As shown in Fig.~\ref{fig:couette}, a reasonable agreement in the predicted profile of number density, bulk velocity, temperature, and heat flux for the Gamma-Boltzmann model compared with the benchmarks is obtained. Furthermore, we have studied the cost of the Gamma-Boltzmann particle scheme for the Couette flow at different densities compared to the benchmarks. As shown in Fig.~\ref{fig:couette_cost}, similar to FP model and unlike DSMC solution, the cost of the new scheme does not scale with density nor temperature. Hence, the Gamma-Boltzmann model can provide an efficient alternative approximation to the Boltzmann equation for non-equilibrium fluid flows at small Knudsen numbers.

\begin{figure}
  \centering
  \begin{tabular}{cc}
  (a) $N_{p/\mathrm{cell}}=100$ & (b) $N_{p/\mathrm{cell}}=200$
\\
\includegraphics[scale=1.0]{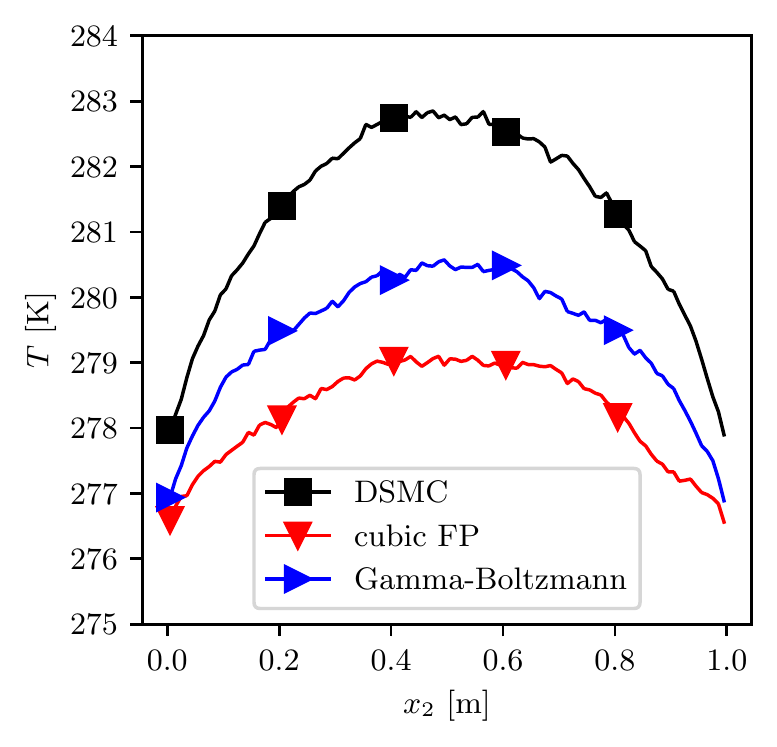}&
\includegraphics[scale=1.0]{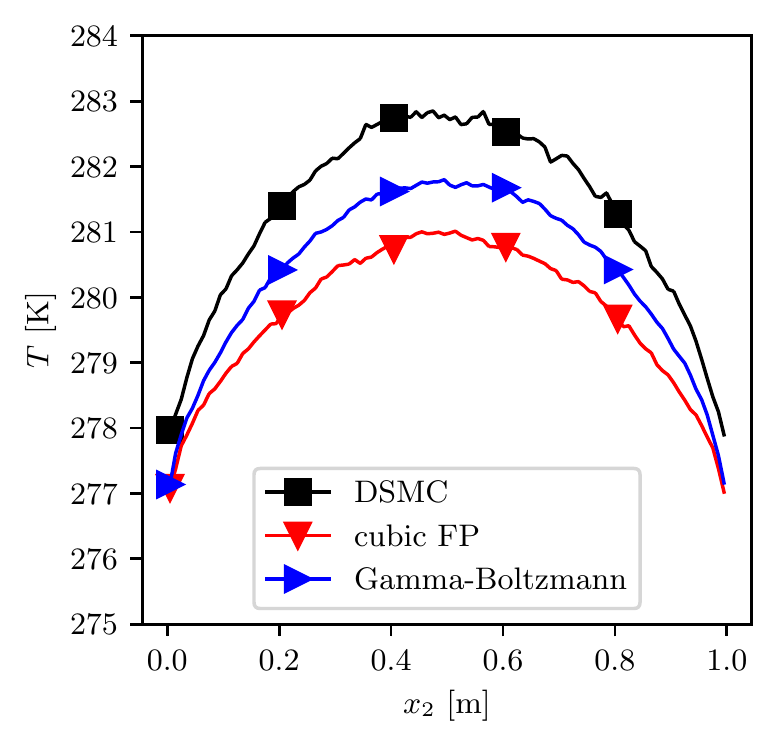}
\\
(c) $N_{p/\mathrm{cell}}=400$ & (d) $N_{p/\mathrm{cell}}=1000$
\\
\includegraphics[scale=1.0]{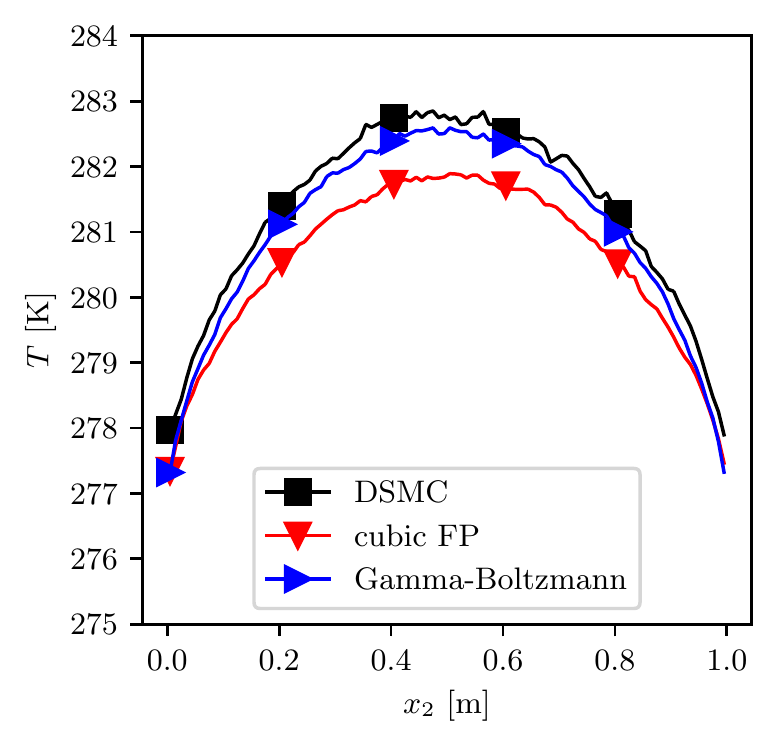}
&
\includegraphics[scale=1.0]{Figures/couette/T_Np1000.pdf}
  \end{tabular}
  \caption{The temperature profile of Couette flow obtained from simulation of jump-diffusion process (blue), cubic FP model (red) using initially $N_{p/\mathrm{cell}}=100,200,400,$ and $1000$ particles per cells. As the reference, the temperature profile obtained from the DSMC solution (black) with initially $N_{p/\mathrm{cell}}=2000$ particles per cell is shown in all figures as the reference solution.}
  \label{fig:analysis_Np}
\end{figure}

\ \\
\noindent Furthermore, we  compare the solution obtained from the devised jump-diffusion process against the cubic FP model in the limit of low number of particles. Here, we simulate the Couette flow using initially $N_{p/\mathrm{cell}}=100,\ 200,\ 400,\ \mathrm{and}\ 1000$ particles per cell. Once stationary state is achieved ($5000$ steps), we average the moments in time until the noise level in the profile of temperature is below $5\%$. This analysis allows us to investigate the error in each model due to lack of particles. As shown in Fig.~\ref{fig:analysis_Np}, the devised jump-diffusion process provides a more accurate solution compared to the cubic FP when less particles are available.
This can be explained by the fact that the Gamma-Boltzmann model requires an estimate of lower order moments (third order) compared to cubic FP model (fifth order). Therefore, the jump-diffusion process is less prone to error due to statistical noise.

\FloatBarrier
\subsection{Lid-driven cavity}
\label{sec:lid-driven}
\begin{figure}
\thisfloatpagestyle{empty}
  \centering
  \begin{tabular}{cc}
\includegraphics[scale=0.95]{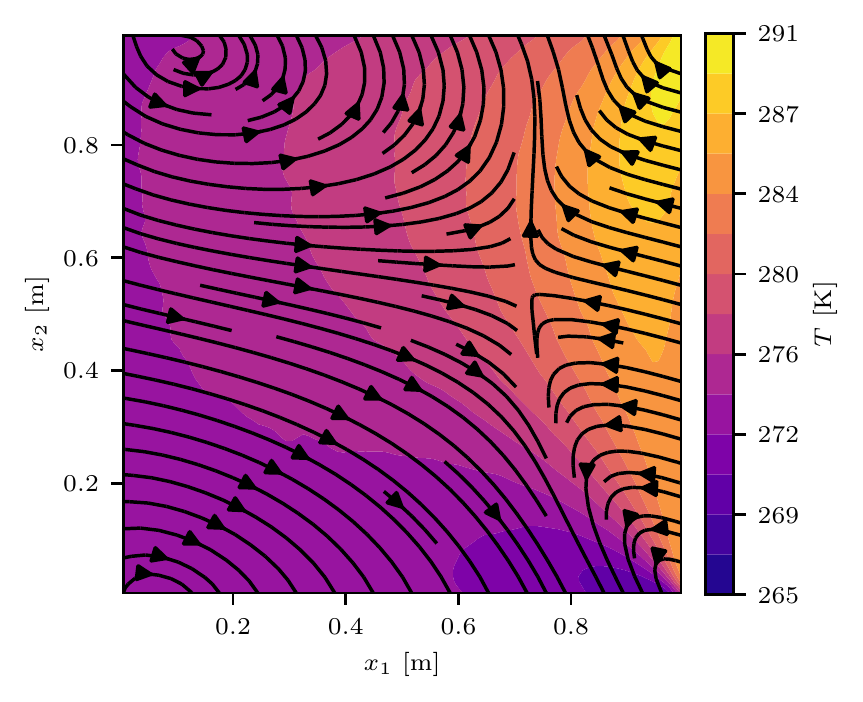}&
\includegraphics[scale=0.95]{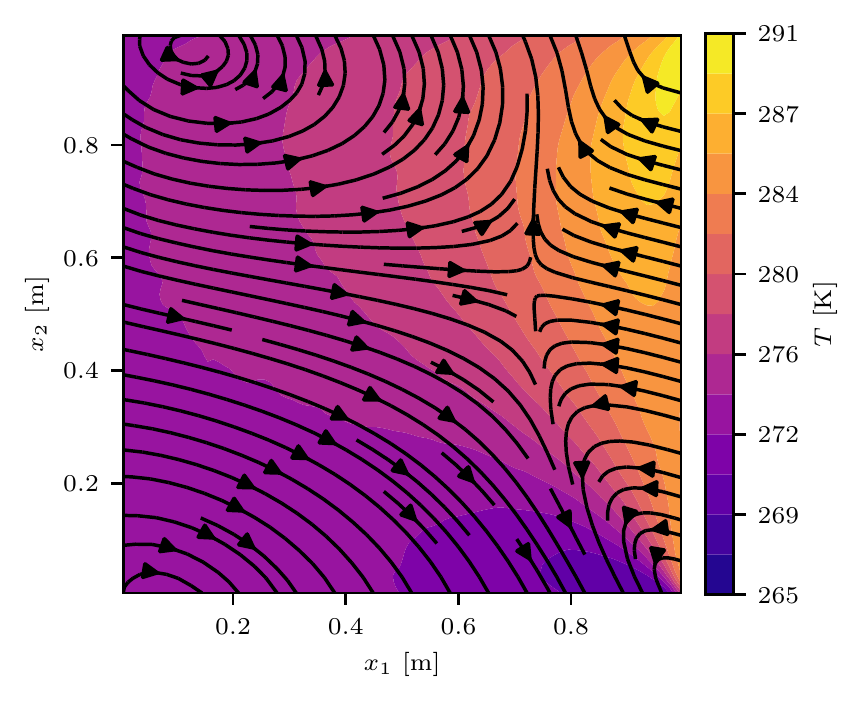}\\
(a) & (b)\\
\includegraphics[scale=0.95]{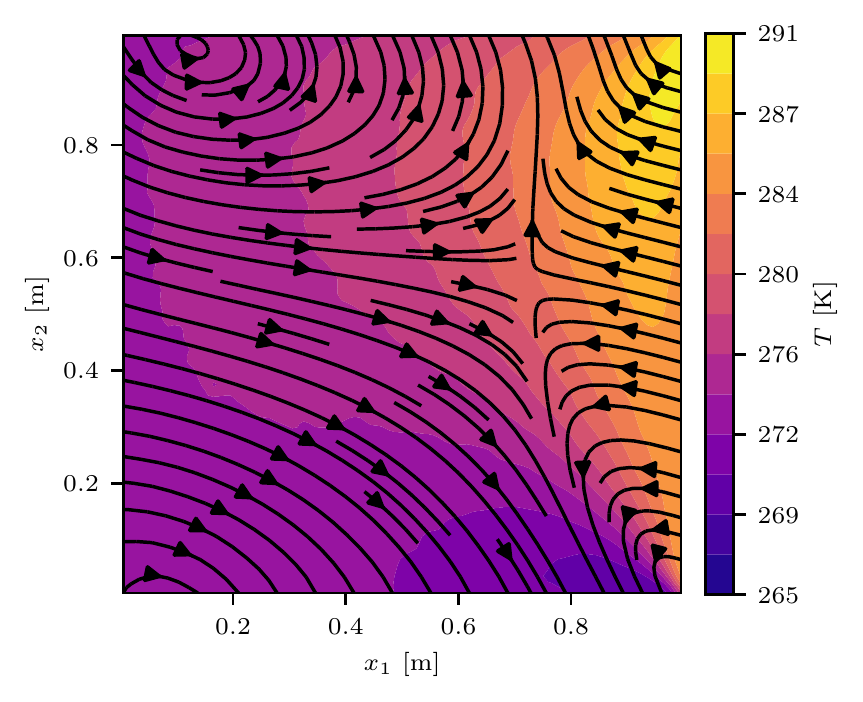} & \includegraphics[scale=1]{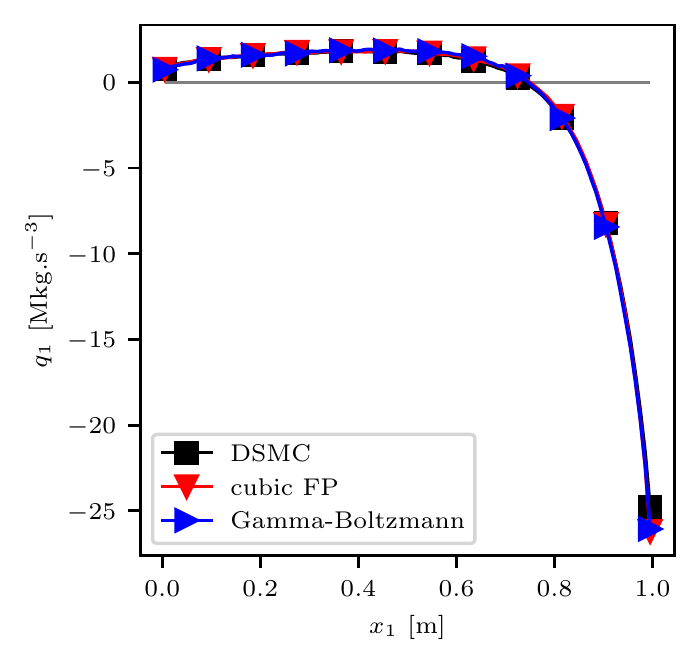}
\\
(c) & (d)
\\
\includegraphics[scale=0.95]{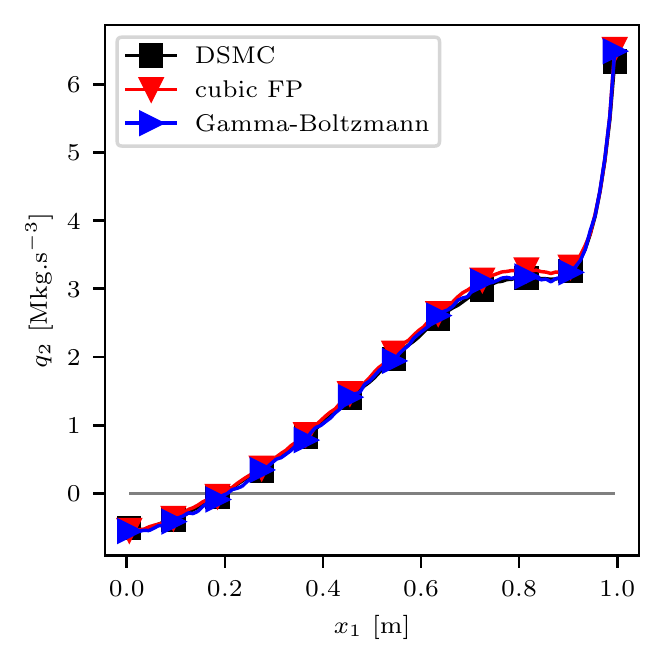} & \includegraphics[scale=1]{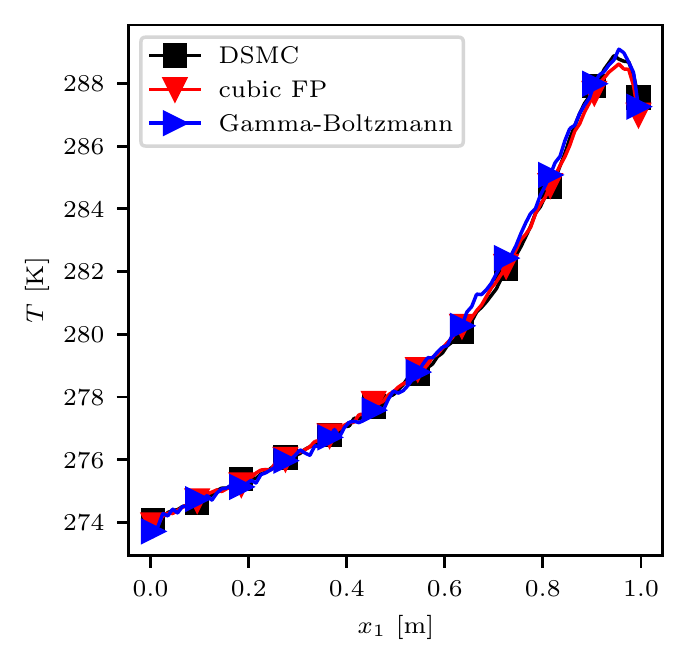}
\\
(e) & (f)
  \end{tabular}
  \caption{Temperature contours overlaid by the heat flux curves in the lid-driven cavity flow with $u_\mathrm{nw}=(150,0,0)^T\ \mathrm{m.s^{-1}}$ at $\mathrm{Kn}=0.17$ obtained from simulation of (a) DSMC, (b) cubic FP and (c) Gamma-Boltzmann. The heat fluxes and temperature on $x_2=4L/5$  line is plotted against the benchmark in (d), (e) and (f).}
  \label{fig:lid_driven}
\end{figure}

One of the classical fluid problems with a clear non-equilibrium effect is the lid-driven cavity at high Knudsen numbers. Consider a particle system inside $\Omega=[0,L]^2$ where all the walls are taken to be constant and thermal with temperature  of $T_{w}=273\ \mathrm{K}$, except for the northern wall which moves with the velocity of $\bm{u}_\mathrm{nw}=(150,0,0)^T\ \mathrm{m.s^{-1}}$. The boundary conditions on the walls for the particles leaving the domain are imposed in a similar manner to the one of Couette flow which is explained in \S~\ref{sec:couette}. Here, we initially deploy $N_\mathrm{p/cell}=2000$ particles per cell, discretize the spatial domain $\Omega$ uniformly with $N_\mathrm{cell}=100\times 100$ cells and considered time step size of $\Delta t = 2.08\times 10^{-6}\ \mathrm{s}$. The stationary solution is achieved after $5'000$ steps and the moments are time-averaged for $65'000$ steps. 
\\ \ \\
A comparison of the temperature and heat fluxes obtained from simulation of DSMC, cubic FP, and the Gamma-Boltzmann model is shown in Fig.~\ref{fig:lid_driven}. As expected, we observe the cold-to-hot heat flux as a non-equilibrium effect in all simulation results. Overall, a reasonable agreement between the developed Gamma-Boltzmann model and the benchmarks in the estimation of moments up to heat flux is obtained. 

\FloatBarrier
\section{Conclusion}
 \label{sec:conclusion}
In this work, we devised a jump-diffusion process which approximates the solution of the Boltzmann equation up to heat fluxes. In particular, we adapted the linear Fokker-Planck model by adding a jump process which provides us with an explicit evolution of particle velocity. While the proposed Gamma-Boltzmann particle scheme avoids performing explicit collisions as particles follow independent paths, the computational effort increases near the equilibrium. We tackled this numerical challenge by replacing the exact trajectories with an approximation which provide us with appropriate efficiency in simulations. The devised solution algorithm was tested against the ones obtained from the DSMC and cubic FP model in several test cases, such as the Couette flow and lid-driven cavity. Overall, a reasonable agreement between the Gamma-Boltzmann model and the benchmark has been observed. Furthermore, we observe that the devised Gamma-Boltzmann model gives a more accurate solution in the noisy settings compared to the cubic FP as it does not require estimation of high order moments in comparison.
\\ \ \\
The specification of particle dynamics in terms of jump-diffusion processes allows for great flexibility. Future work might explore the use of different jump intensity measures to approximate the Boltzmann operator with more accuracy, e.g.\ by matching the evolution of higher order moments and the entropy production. Furthermore, to avoid the approximation made for Gamma-Boltzmann model near equilibrium, a more accurate solution near equilibrium may be achieved by coupling the jump-diffusion process  with the ellipsoidal Fokker-Planck model, where one can switch between both dynamics according to the distance of the gas from equilibrium.

\section*{Acknowledgements}
MS acknowledges the funding provided by the German research foundation (DFG) under the grant number  SA 4199/1-1.

\appendix
\section*{Appendices}

\maketitle
\end{comment}

\section{Evolution of moments for the Gamma process jump operator}

In this section, we study the integral operator $\mathcal{S}^{\text{J}}$ corresponding to the jump process, which is defined as
\begin{align*}
    \mathcal{S}^\text{J} \F(\vv; \vx, t) 
	&= \int \left[\F(\vv-\vc) - \F(\vv)\right]\, \nu^\Gamma(d\vc, \vx,t), 
\end{align*}
with intensity measure
\begin{align}
	\nu^\Gamma(d\vc,\vx,t) & = \sum_i \gamma_i\frac{\exp(-c_i/\lambda_i)}{|c_i|} {\mathds{1}}(\lambda_i c_i >0)\, S_{\R_i}(d\vc). \label{eqn:nugamma_app}
\end{align} 
Our goal is to specify the parameters $\gamma_i$ and $\lambda_i$ such that $\mathcal{S}^\text{J}$ may be used to approximate the Boltzmann collision operator $\mathcal{S}^{\text{Boltz}}$.
To this end, we compute the rates $\int \psi(\vv)\mathcal{S}^\text{J} \F(\vv)\, d\vv$ for the intensity measure $\nu^\Gamma$.

Recall that 
\begin{align}
	\int \psi(\vv)\mathcal{S}^\text{J} \F(\vv; \vx, t)\, d\vv 
	&= \int \int \left[\psi(\vv + \vc) - \psi(\vv)\right] \nu(d\vc, \vx, t) \F(\vv; \vx,t)\, d\vv. \label{eqn:moment-jumps_app}
\end{align} 
We may employ \eqref{eqn:moment-jumps_app} to obtain, for $i=1,2,3,$
\begin{align*}
	\int v_i \mathcal{S}^\text{J} \F(\vv;\vx,t)\, d\vv &= \int \int c_i \nu^\Gamma(d\vc;\vx,t)\F(\vv;\vx,t) \, d\vv = \lambda_i(\vx,t) \gamma_i(\vx,t) \rho(\vx, t), 
\end{align*}
which is non-zero. 
Hence, to ensure conservation of momentum, we need to introduce an additional linear term $-\lambda_i\gamma_i$ to the drift $\vmu$.
That is, we consider the operator
\begin{align*}
	\mathcal{S}^\Gamma \F(\vv) &= \mathcal{S}^{J} \F(\vv) + \sum_i\frac{\partial}{\partial v_i}\left( \lambda_i \gamma_i \F(\vv) \right),
\end{align*}
such that
\begin{align*}
	\int v_i \mathcal{S}^\Gamma \F(\vv)\, d\vv = 0. 
\end{align*}
Furthermore, applying \eqref{eqn:moment-jumps_app} to the function $\psi(\vv)=1$, we obtain
\begin{align*}
	\int \mathcal{S}^\Gamma \F(\vv)\, d\vv = 0.
\end{align*}
That is, the operator $\mathcal{S}^\Gamma$ satisfies conservation of mass and momentum.
Furthermore, the second moments may be determined as
\begin{align*}
	&\int w_i w_j \mathcal{S}^\Gamma \F(\vv;\vx,t)\, d\vv \\
	&=  \int \int \left[c_i c_j +  w_i c_j + w_j c_i\right] \nu^\Gamma(d\vc,\vx,t)\F(\vv;\vx,t) \, d\vv - \lambda_i\gamma_i\int w_j \F(\vv)\, d\vv - \lambda_j\gamma_j\int w_i \F(\vv)\, d\vv \\
	&= \int \int c_i c_j  \nu^\Gamma(d\vc;\vx,t)\F(\vv;\vx,t) \, d\vv,
\end{align*}
because $\int w_i \F(\vv)\, d\vv =0$, and $\int c_i \nu^\Gamma(d\vc, x, t)=\lambda_i\gamma_i$.
Since $\nu^\Gamma$ is only supported on the axis $\R_i$, we find that 
\begin{align*}
	\int c_j c_i  \nu^\Gamma(d\vc, \vx,t) &= \delta_{ij} \gamma_i \int_0^\infty \frac{c_i^2}{|c_i|} \exp(-c_i/\lambda_i)\, dc_i
	= \delta_{ij} \gamma_i \lambda_i^2 , \quad i,j=1,2,3,
\end{align*}
so that $\int w_iw_j \mathcal{S}^\Gamma \F(\vv; \vx,t)\, d\vv = \delta_{ij} \rho \gamma_i \lambda_i^2$.
In particular, the operator $\mathcal{S}^\Gamma$ does not conserve internal energy on its own. 
This can be corrected by choosing a suitable mean-reverting drift term $\vmu$, as demonstrated in the next section.

Regarding the heat flux, we observe that
\begin{align*}
	&\int w_i \sum_k w_k w_k \mathcal{S}^\Gamma \F(\vv)\, d\vv \\
	&= \int w_i \sum_k w_k w_k \mathcal{S}^\text{J} \F(\vv)\, d\vv - \int \sum_k \left[ 2 w_i w_k \lambda_k \gamma_k + w_k w_k \lambda_i \gamma_i \right] \F(\vv)\, d\vv, \\
	\textrm{where\ \ \ }&\int w_i \sum_k w_k w_k \mathcal{S}^\text{J} \F(\vv)\, d\vv \\
	&= \int \int \sum_k \left[ (w_i+c_i)(w_k+c_k)(w_k+c_k) - w_i w_k w_k \right] \nu^\Gamma(d\vc)\, \F(\vv)\, d\vv \\
	&= \int \int \sum_k\left[ c_i c_k c_k  {+ 2 c_iw_k c_k + w_i c_k c_k} + c_iw_k w_k + 2w_i w_k c_k  \right] \nu^\Gamma(d\vc)\, \F(\vv)\, d\vv \\
	&= \int \int  c_i \sum_k c_k c_k  \nu^\Gamma(d\vc)\, \F(\vv)\, d\vv +  \int \sum_k\left[ 2w_i w_k \lambda_k \gamma_k + \lambda_i\gamma_i w_k w_k \right]\, \F(\vv)\, d\vv.
\end{align*}
In the last step, we used that
\begin{align*}
    \int \int \sum_k c_i w_k c_k   \nu^\Gamma(d\vc)\, \F(\vv)\, d\vv  
    &= \sum_k\int c_i c_k\int   w_k     \F(\vv)\, d\vv  \, \nu^\Gamma(d\vc) =0, \\
    \int \int \sum_k w_i c_k c_k  \nu^\Gamma(d\vc)\, \F(\vv)\, d\vv =0,
\end{align*}
because $\int w\F(\vv)\, d\vv=0$.
Since $\nu^\Gamma$ is only supported on the axes, we find that
\begin{align*}
	\int w_i \sum_k w_k w_k \mathcal{S}^\Gamma \F(\vv)\, d\vv 
	&= \int \int  c_i \sum_k c_k c_k  \nu^\Gamma(d\vc)\, \F(\vv)\, d\vv \\
	&= \rho  \int  c_i \sum_k c_k c_k \delta_{ik}  \nu^\Gamma(d\vc) \\
	&= \rho  \gamma_i \lambda_i^3 \int_0^\infty \frac{c_i^3}{|c_i|} \exp(-c_i)\, dc_i  \\
	&= 2\rho \gamma_i \lambda_i^3.  
\end{align*}

To summarize, we obtain
\begin{align*}
    \int 1\cdot \mathcal{S}^\Gamma \F(\vv;\vx,t)\, d\vv &= 0,\\ 
	\int v_i \mathcal{S}^\Gamma \F(\vv;\vx,t)\, d\vv &= 0, \\
	\int w_iw_j \mathcal{S}^\Gamma \F(\vv; \vx,t)\, d\vv 
	&= \delta_{ij} \rho(\vx,t)\, \gamma_i(\vx,t)\, \lambda_i(\vx,t)^2, \\
	\int w_i \sum_k w_k w_k \mathcal{S}^\Gamma \F(\vv;\vx,t)\, d\vv 
	&= 2\rho(\vx,t)\, \gamma_i(\vx,t)\, \lambda_i(\vx,t)^3.
\end{align*}
For the jump operator $\mathcal{S}^J$ without mean correction, this implies
\begin{align*}
    \int 1\cdot \mathcal{S}^\Gamma \F(\vv;\vx,t)\, d\vv &= 0,\\
    \int v_i \mathcal{S}^J \F(\vv;\vx,t)\, d\vv &= \rho(\vx,t)\, \lambda_i(\vx,t)\, \gamma_i(\vx,t) \\
    \int w_i  w_j \mathcal{S}^J \F(\vv;\vx,t)\, d\vv &= \delta_{ij} \rho(\vx,t)\, \lambda_i(\vx,t)^2\, \gamma_i(\vx,t) \\
    \int w_i \sum_k   w_k w_k \mathcal{S}^J \F(\vv;\vx,t)\, d\vv &= 2\rho(\vx,t)\, \lambda_i(\vx,t)^3\, \gamma_i(\vx,t)  \\
    &\quad + \int \sum_k (2w_i w_k \lambda_k \gamma_k + \lambda_i \gamma_i w_k w_k) \F(\vv;\vx,t)\, d\vv.
\end{align*}

\section{Fixing the Prandtl number}

In this section, we devise the full jump-diffusion model in velocity space 
\begin{align}
	\begin{split}
	d \vv(t) &= \vmu dt + \vF(\vx(t), t)dt + \mSigma^\frac{1}{2} d\vW(t) + \int \vc(\vz, \vx(t),t)\, N(d\vz, dt)\label{eqn:JD_rep}
	\end{split}
\end{align}
by setting
\begin{flalign*}
	\mu_i &= -a (v_i-u_i) - \lambda_i \gamma_i,\\
	\Sigma_{ij} &=  b \delta_{ij} \frac{\sum_{k} p_{kk}}{3} .
\end{flalign*}
Moreover, we choose $\vc(\vz, \vx,t)$ and the underlying intensity measure $\nu(dz)$ such that the local jump intensity measure is given by
\begin{flalign*}
    \nu(d\vc, \vx, t) &= \nu^\Gamma(d\vc,x,t).
\end{flalign*}
where $\nu^\Gamma(dz)$ is as in \eqref{eqn:nugamma_app}, and $\gamma_i\geq 0$ and $\lambda_i\in\R\setminus\{0\}$ are functions of location $x$ and the density $\F$.
Note that the specific choice of $\nu(d\vz)$ and $\vc(\vz,\vx,t)$ is irrelevant, as long as they correspond to the local jump intensity measure $\nu(d\vc,\vx,t)$.

The values $a,b$, and the specific form of $\lambda_i, \gamma_i$, need yet to be specified.
The corresponding collision operator $\mathcal{S}^\text{JD}$ is given by
\begin{align*}
	\mathcal{S}^\text{JD}\F(\vv; \vx,t) &= \mathcal{S}^\text{FP}\F(\vv; \vx,t) + \mathcal{S}^\Gamma\F(\vv; \vx,t) \\ 
	\text{where}\ \ \ 
	\mathcal{S}^\text{FP}\F(\vv; \vx,t)&= \sum_{i} \frac{\partial}{\partial v_i} \left(\,a(v_i-u_i)\, \F(\vv; \vx,t) \right) +\frac{1}{2}\sum_{i,j} \frac{\partial^2}{\partial v_i \partial v_j}\left(\Sigma_{ij} \F(\vv; \vx,t) \right),
	\\
    \mathcal{S}^\Gamma\F(\vv; \vx,t) &=  \int \left[\F(\vv-\vz) - \F(\vv)\right]\, \nu(d\vz;\vv, \vx,t) + \sum_i \lambda_i \gamma_i \frac{\partial}{\partial v_i} \F(\vv).
\end{align*}
Just like $\mathcal{S}^\text{FP}$ and $\mathcal{S}^\Gamma$, the operator $\mathcal{S}^\text{JD}$ conserves mass and momentum for any value of $a$ and $b$, i.e.\
\begin{align*}
    \int 1\cdot \mathcal{S}^\text{JD}\F(\vv; \vx,t)\, d\vv &=0, \\
    \int v_i \cdot \mathcal{S}^\text{JD}\F(\vv; \vx,t)\, d\vv &=0. 
\end{align*}
In the sequel, we will omit the dependency on $(\vx,t)$.
From our previous derivations, we find that
\begin{align*}
	\sum_k \int w_k w_k \mathcal{S}^\text{JD} \F(\vv)\, d\vv 
	&= (b-2a) 	\sum_k p_{kk} + \rho 	\sum_k \gamma_k \lambda_k^{2}, \\
	\int w_{\langle i} w_{j\rangle} \mathcal{S}^\text{JD} \F(\vv)\, d\vv 
	&= -2a p_{\langle ij\rangle} + \delta_{ij} \rho \left[\gamma_i \lambda_i^2 - \sum_k \frac{\gamma_k\lambda_k\lambda_k}{3}\right],\\
	\frac{1}{2}\int w_i \sum_k w_k w_k \mathcal{S}^\text{JD}\F(\vv)\, d\vv
	&= -3a q_i + \rho \gamma_i\lambda_i^3.
\end{align*}
For some $0\leq c \leq 2a$, we suggest to choose $\gamma_i$ and $\lambda_i$ such that
\begin{align*}
	\gamma_i \lambda_i^2 &= \frac{c}{3} \frac{\sum_k p_{kk}}{\rho},\\
	\gamma_i \lambda_i^3 &= \frac{5a}{3}  \frac{q_i}{\rho},
\end{align*}
that is 
\begin{align*}
    \lambda_i = \frac{5a}{c} \frac{q_i}{\sum_k p_{kk}}, 
    \qquad
    \gamma_i = \frac{c^3}{75 a^2} \frac{1}{\rho} \frac{(\sum_k p_{kk})^3}{q_i^2}.
\end{align*}
This choice yields 
\begin{align}
	\begin{split}
	\int \sum_k w_k w_k \mathcal{S}^\text{JD} \F(\vv)\, d\vv 
	&= (b+c-2a) \sum_k p_{kk} , \\
	\int w_{\langle i} w_{j\rangle} \mathcal{S}^\text{JD} \F(\vv)\, d\vv 
	&= -2a p_{\langle ij\rangle} ,\\
	\frac{1}{2}\int w_i \sum_k w_k w_k \mathcal{S}^\text{JD}\F(\vv)\, d\vv 
	&= -\frac{4}{3}a q_i.\label{eqn:JD-rates}
	\end{split} 
\end{align}
Hence, conservation of energy is satisfied if $b+c=2a$.
Furthermore, this model gives the correct Prandtl number $\tau_\text{JD} =\tau_\text{Boltz} = \frac{2}{3}$.
The choice of $c\in(0,2a]$ is a remaining degree of freedom.

\section{Particle dynamics near equilibrium}

If $q_i=0$, the values of $\gamma_i$ and $\lambda_i$ are not well-defined. 
To extend our model to this regime, we consider the limit as $q_i\to 0$.
In this situation, $\lambda_i\to 0$ and $\gamma_i\to\infty$ such that the variance $\gamma_i\lambda_i^2$ of the jump process remains constant. 
In particular, the jumps become smaller but also more frequent.
It turns out that a type of central limit theorem applies, such that in the limit, the jump process becomes a continuous Gaussian movement.
To make this precise, we observe that
\begin{align*}
    \mathcal{S}^{\Gamma} \F(\vv) = \int \left[\F(\vv-\vc) - \F(\vv) - \sum_i c_i \frac{\partial}{\partial v_i} \F(\vv) \right]\, \nu^\Gamma(d\vv),
\end{align*}
because $\int c_i \nu^\Gamma(d\vv) = \lambda_i\gamma_i$.
Now recall from \eqref{eqn:nugamma_app} that $\nu^\Gamma$ is the sum of three measures supported on the axes $\mathbb{R}_i$.
With some abuse of notation, we write $\vv+c_i$ to mean that $c_i$ is added to the $i$-th component of the vector $\vv$. 
Then the jump operator may be written as
\begin{align*}
    \mathcal{S}^{\Gamma} \F(\vv) &= \sum_i \mathcal{S}^\Gamma_i \F(\vv), \\
    \mathcal{S}^\Gamma_i \F(\vv) &= \int_0^\infty \left[\F(\vv-\lambda_i c_i) -\F(\vv)\right] \gamma_i \frac{\exp(-c_i)}{c_i}\, dc_i.
\end{align*}
If the third order derivatives of $\F$ are bounded, then a Taylor expansion gives
\begin{align*}
     \mathcal{S}^{\Gamma}_i \F(\vv) 
     &=\frac{1}{2}\int_0^\infty c_i^2 \frac{\partial^2}{\partial v_i^2}\F(\vv) \, \gamma_i \frac{\exp(-c_i)}{c_i}\, dc_i 
     + \mathcal{O}\left(\int_0^\infty |\lambda_i c_i|^3 \, \gamma_i \frac{\exp(-c_i)}{c_i}\, dc_i\right) \\
     &= (\lambda_i^2 \gamma_i)\frac{1}{2}  \frac{\partial^2}{\partial v_i^2}\F(\vv) +  \mathcal{O}(|\lambda_i^3 \gamma_i|).
\end{align*}
As $q_i\to 0$, the second term vanishes, so that $\mathcal{S}^\Gamma_i$ converges towards a diffusion operator. 
That is, as $q_i\to 0$, the jumps in dimension $i$ increasingly resemble a continuous Gaussian movement with diffusion coefficient $\lambda_i^2\gamma_i$. 

To incorporate this limiting behaviour in the definition of the jump diffusion model \eqref{eqn:JD_rep}, we modify the equation by setting
\begin{align*}
    \Sigma_{ij} &= b\delta_{ij} \frac{\sum_k p_{kk}}{3} + \delta_{ij} \frac{c}{\rho} \frac{\sum_k p_{kk}}{3} \mathds{1}_{q_i=0}, \\
    \gamma_i &= \frac{c^3}{75 a^2} \frac{1}{\rho} \frac{(\sum_k p_{kk})^3}{q_i^2} \mathds{1}_{q_i\neq 0}.
\end{align*}
Thus, the particle evolution is well defined for all cases, and admits the correct Prandtl number, for any choice $a,b,c$, such that $b+c=2a$.

If $q_i=0$ for all $i$, then $\mathcal{S}^{\text{JD}}$ reduces to the Fokker-Planck operator $\mathcal{S}^\text{FP}$ as studied by \cite{Jenny2010}.
Hence, we also find that $\mathcal{S}^\text{JD} \F_\text{eq}=0$, where $\F_\text{eq}$ denotes the Maxwellian equilibrium distribution.
On the other hand, if $\mathcal{S}^\text{JD} \F= 0$, then we find that $q_i=0$ from the moment evolutions. Thus, $\mathcal{S}^\text{JD} \F= \mathcal{S}^\text{FP} \F= 0$, which implies $\F = \F_\text{eq}$.



\bibliography{levyBoltzmann,levyBoltzmann2}
\bibliographystyle{apalike}

\end{document}


\maketitle

This attachment serves as a brief, but rigorous, introduction to stochastic processes with jumps, with a special focus on jump diffusions. 
After describing the necessary theory to define jump diffusion processes, we in particular present the evolution equations for the marginal moments, and discuss the relation to the Kolmogorov-forward equation which describes the evolution of the corresponding probability density function.

\section{Basics}\label{basics}

A stochastic process \((X_t)_{t\geq 0}\), is a collection of random vectors, which is in our case indexed by time \(t\). 
Formally, we endow the scenario-set \(\Omega\) with a \(\sigma\)-algebra \(\mathcal{F}\) and a probability measure \(P\) on \((\Omega,\mathcal{F})\), such that each \(X_t:\Omega\to\R^d\) is a measurable mapping. 
If we fix some \(\omega\in\Omega\), the mapping \(t\mapsto X_t(\omega)\) is called the trajectory of \(X_t\).
Hence, a stochastic process may be regarded as a random function. 
The shape of these trajectories depends on the probabilistic properties of the process \((X_t)_{t\geq 0}\).
There exist examples of stochastic processes whose trajectories are discontinuous, continuous, or very smooth. 
In the sequel, the dependence on \(\omega\) will be implicit.

A minimal assumption imposed in most of the literature on stochastic processes is that the trajectories are right continuous and admit finite left-hand limits, that is for each \(t> 0\), the left-hand limit \(X_{t-}:=\lim_{s\uparrow t} X_s\) exists, and for all \(t\geq 0\), the right-hand limit satisfies \(\lim_{s\downarrow t} X_s = X_t\). 
This property is often referred to by the french acronym \emph{càdlàg} (\emph{continue à droite, limite à gauche}). 
The trajectories of a càdlàg process only admit jump-type discontinuities, but no singularities. 
We denote the jump size at time \(t\) by \(\Delta X_t = X_t - X_{t-}\), which is a random variable, i.e.\
\(\Delta X_t = \Delta X_t(\omega)\). 
An important property of càdlàg functions is that, on any finite interval, they admit only finitely many jumps of size larger than some arbitrary \(\epsilon>0\), i.e.\
\(\sum_{t\in[0,T]} \mathds{1}(|\Delta X_t(\omega)|>\epsilon) < \infty\) \citep[Thm.\ 2.9.2]{applebaum2009levy}.
As a consequence, the set of jump times \(\{t\geq 0 : \Delta X_t \neq 0 \}\) is countable.
Hence, sums of the form \(\sum_{t\geq 0} f(\Delta X_t, t) \) are sensible as countable sums for any function $f$ such that \(f(0,t)=0\) , although the summation range is formally uncountable. 
In this appendix, we explain the construction and calculus of a special class of discontinuous processes, namely jump diffusions.

\section{Constructive definition of jump processes}\label{constructive-definition-of-jump-processes}

\begin{figure}[tb]
	\centering
	\includegraphics[width=0.8\textwidth]{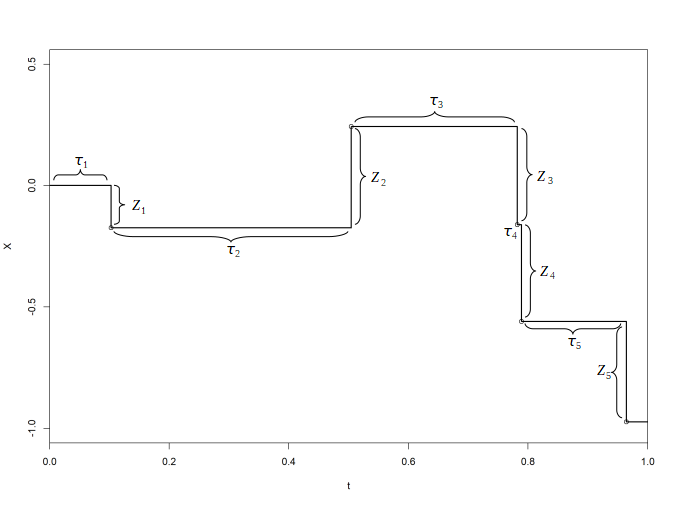}
	\caption{Sample path of a compound Poisson process, with uniformly distributed jump sizes $Z_i \sim U(-1,1)$.}
	\label{fig:compount-poisson}
\end{figure}

A simple jump process may be constructed as follows. 
Let \(\tau_i, i\in\N\), be a sequence of independent, exponentially distributed random variables with rate parameter \(\lambda_0>0\), \(\tau_i \sim \text{Exp}(\lambda_0)\), and let \(Z_i, i\in\N\), be a sequence of independent random vectors with probability distribution \(Q\). 
Define the process \(X_t\) as
\begin{align}
	X_t 
	= \sum_{i=1}^{\infty} Z_i \cdot \mathds{1}\left(t \leq \sum_{j=1}^i \tau_j\right) 
	= \sum_{i=1}^{N_t} Z_i,\qquad N_t = \inf \left\{k\in\N_0 : t\geq \sum_{i=1}^k \tau_i \right\}. \label{eqn:compound-poisson}
\end{align}
That is, after the \(i\)-th jump, we wait for a random time \(\tau_{i+1}\sim \text{Exp}(\lambda_0)\) until the next jump, and then sample the jump size \(Z_{i+1}\) according to the distribution \(Q\). 
In other words, the jump times are \(T_i = \sum_{j=1}^i \tau_j\), and the jump sizes are \(\Delta X_{T_i} = Z_i\). 
The process defined via \(\eqref{eqn:compound-poisson}\) is a so-called compound Poisson process. 
An example for the univariate case \(d=1\) is depicted in Figure \ref{fig:compount-poisson}.

The exponential distribution is chosen for the waiting times because of
its memoryless property, i.e.\ \(P(\tau_i>t+s|\tau_i \geq t) = P(\tau_i>s)\) for all \(t,s\geq 0\). 
That is, having already waited for time \(t\) does not change the waiting time, i.e. the remaining waiting time may not be predicted. 
In other words, the jumps of \(X_t\) occur suddenly. 
A consequence of the memoryless property is that \(X_t\) is a special instance of a Lévy process. 
Lévy processes constitute a broad class of stochastic processes defined as follows.

\begin{definition}
	A stochastic process \((X_t)_{t\geq 0}\) is a Lévy process if it satisfies:
	\begin{itemize}
		\item
		\(X_0=0\) almost surely.
		\item
		The increments are independent and stationary, i.e. \(X_{t+s}-X_t\), \(s\geq 0\), is independent of \(\{X_r, r\in[0,t]\}\), and \(X_{t+s}-X_t\) has the same distribution as \(X_s\).
		\item
		The process is stochastically continuous, i.e.
		\(P(|X_{t+s}-X_t|>\epsilon) \overset{s\to 0}{\longrightarrow} 0\) for any \(t\geq 0\) and any \(\epsilon>0\).
	\end{itemize}
\end{definition}

Another prominent example of a Lévy process is the Brownian motion \((B_t)_{t\geq 0}\), where \(B_{t+s}-B_t \sim \mathcal{N}(0,s\sigma^2)\) for \(s\geq 0\) and some \(\sigma^2>0\). 
It is tempting to try to define a process by specifying any arbitrary distribution \(Q_s\) for the increments \(X_{t+s}-X_t\), and then construct \(X_t\) by summing up the independent increments. 
This is, however, in general not possible, because the family of probability measures \((Q_s)_{s>0}\) needs to be
consistent with the property of independent and stationary increments for all \(s\geq 0\). 
In particular, 
\(X_{t+s_1+s_2}-X_t = (X_{t+s_1+s_2}-X_{t+s_1}) + (X_{t+s_1}-X_t)\), 
where both terms are independent and distributed as \(Q_{s_1}\) and \(Q_{s_2}\). 
Hence, the family of measures needs to satisfy \(Q_{s_1}\ast Q_{s_2} = Q_{s_1+s_2}\), where \(\ast\) denotes
convolution of measures. 
The set of distributions which may consistently be chosen as \(Q_s\) is the class of so called infinitely divisible distributions. 
This class contains, for example, Gaussian and Gamma distributions, but not the uniform distribution.

\section{Integral representation}\label{integral-representation}

The trajectory of the compound Poisson process \(\eqref{eqn:compound-poisson}\) may be completely described by the sequences of the jump times \(T_i = \sum_{j=1}^i \tau_j, i\in\N\), and the jump sizes \((Z_i)_{i\in\N}\). 
An equivalent representation is via the measure \(N=N(dt, dz)\) on \([0,\infty) \times \R^d\), given by 
\(N = \sum_{i=1}^\infty \delta_{(T_i, Z_i)}\). 
Here, \(\delta_{(t,z)}\) denotes the point measure, i.e.\ the delta functional, at \((t,z)\in[0,\infty)\times \R^d\), and the dependence on \(\omega\in\Omega\) is implicit, i.e.\ \(N(dt,dz)=N(dt,dz,\omega)\). 
The compound Poisson process may then be recovered by integrating
\begin{align}
	X_t = \int_0^t \int_{\R^d} z \, N(ds,dz) = \sum_{i=1}^\infty Z_i \, \mathds{1}(T_i \leq t). \label{eqn:jump-integral}
\end{align}
While this formulation might seem unnecessarily complicated at first sight, it allows for an elegant generalization of the compound Poisson process. Since \(N(dt,dz)\) is a measure, it may be evaluated for any Borel set \(A\subset [0,\infty)\times \R^d\). 
It can be shown that for any such \(A\), the measure \(N(A) = \int_A N(dt,dz)\) has a Poisson distribution with parameter \(\lambda(A)\geq 0\), i.e.\
\(P(N(A)=k) = \frac{\lambda(A)^k e^{-\lambda(A)}}{k!}\), \(k\in\N_0\).
In particular, 
\(\mathbb{E} (N(A)) = \int_\Omega \left(\int_A N(dt,dz)\right)P(d\omega)\), where $P$ is the underlying probability measure.
Moreover, if \(A\cap \tilde{A}=\emptyset\), then \(N(A)\) and \(N(\tilde{A})\) are independent; see e.g.\ \cite[Thm.\ 19.2]{sato1999levy}.

For a rectangular set \(A=[t_1, t_2]\times B\), the measure \(N(A)\) counts the number of jumps of size \(z\in B\) occurring at a time \(t\in[t_1,t_2]\), and \(\lambda(A)\) is the expected number of these jumps. 
Since \(A\mapsto N(A)\) is a measure, a so-called a \emph{Poisson random measure}, it can be shown that its expectation \(A\mapsto\lambda(A)\) is a measure as well, the so-called \emph{intensity measure}. 
For the compound Poisson process, we have \(\lambda(dt, dz) = dt\, Q(dz)\), i.e. \(\lambda\) is the product of the Lebesgue measure on \([0,\infty)\) and a finite measure \(Q\). 
In particular, the rate of jumps is $\lambda_0 = Q(\R^d)$, and the distribution of the jump sizes \(Z_i\) is given by the normalized probability measure $Q/Q(\R^d)$.

It is also possible to start with the intensity measure \(\lambda\) and to construct a corresponding Poisson random measure \(N\), see e.g.\ \cite[Sec.\ 0.5]{bertoin1996}. 
That is, \(A\mapsto N(A)=N(A,\omega)\) is a measure and for any Borel set \(A\subset [0,\infty)\times \R^d\), the random variable \(N(A)\) has a Poisson distribution with parameter \(\lambda(A)\), and $N(A)$ and $N(\tilde{A})$ are independent for $A\cap\tilde{A}=\emptyset$. 
Since \(N\) is an integer-valued measure, it may be represented as a sum of point measures, \(N = \sum_{i=1}^\infty \delta_{(T_i, Z_i)}\) for random variables \(T_i=T_i(\omega)\in[0,\infty)\), and random vectors \(Z_i = Z_i(\omega)\in\R^d\). 
In contrast to the special case of the compound Poisson process described above, the \(Z_i\) are not necessarily identically distributed, but nevertheless independent. 
Using this representation of the Poisson random measure, one may try to construct a process \(X_t=\int_0^t \int_{\R^d} z\, N(ds,dz)\) as in \(\eqref{eqn:jump-integral}\) by integrating the Poisson measure. 

The compound Poisson process has finitely many jumps, hence the sum in \eqref{eqn:jump-integral} is always finite so that existence of the integral is trivial. 
If, on the other hand, \(N\) is an infinite measure, the sum is no longer finite. 
To handle this case of inifnitely many jumps, an alternative integrability condition is that the series are absolutely summable in expectation, that is
\begin{align}
	\mathbb{E} \left[\sum_{i=1}^\infty \|Z_i\|\, \mathds{1}(T_i\leq t) \right] 
	= \mathbb{E}\left[\int_0^t \int_{\R^d} \|z\| \, N(ds,dz)\right]
	= \int_0^t \int_{\R^d} \|z\| \, \lambda(ds,dz) \overset{!}{<} \infty. \label{eqn:finite-variation}
\end{align}
Condition \(\eqref{eqn:finite-variation}\) also covers cases where \(\lambda\) is an infinite measure, i.e.\ the process has infinitely many jumps.
For the integrability to hold, the mass of \(\lambda\) needs to concentrate around \(z=0\). 
That is, the process may have infinitely many jumps, but the jumps are sufficiently small to be summable. 
It is even possible to define an integral w.r.t. \(N\) under the weaker condition
\(\int_0^t \int_{\R^d} \min(1, \|z\|^2) \lambda(ds,dz)<\infty\) 
using an \(L_2(P)\)-isometry, see e.g.\ \cite[Sec.\ 4.2]{applebaum2009levy}. 
However, for the purposes of this paper, condition \eqref{eqn:finite-variation} suffices.

\begin{figure}[tb]
	\centering
	\includegraphics[width=0.49\textwidth]{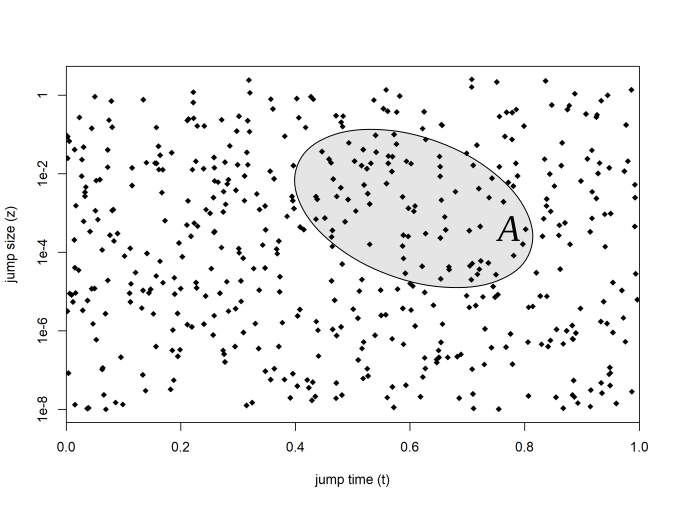}
	\includegraphics[width=0.49\textwidth]{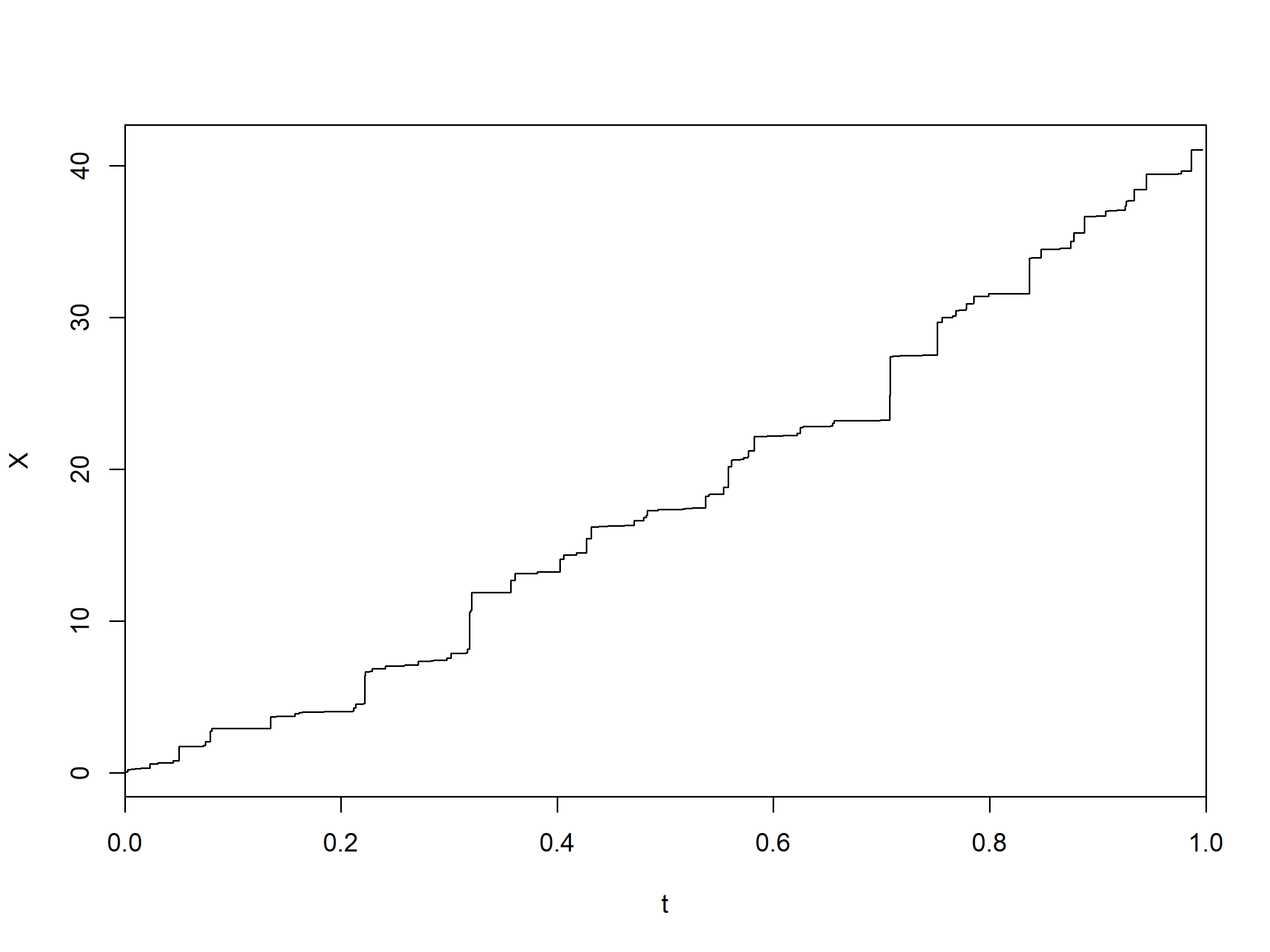}
	\caption{Left: realization of a Poisson random measure with intensity $\lambda(dt,dz)= dt\otimes(20e^{-z}/z)_+ dz$. Right: the corresponding jump process $X_t = \int_0^t \int_{\R^d} z N(ds,dz)$.}
	\label{fig:poisson-measure}
\end{figure}

In Figure \ref{fig:poisson-measure}, we depict the realization of a Poisson random measure $N(dt,dz)$, and the corresponding jump process $X_t = \int_0^t \int_{\R^d} z\, N(ds,dz)$. 
The intensity measure is homogeneous in time, $\lambda(dt,dz) = dt \, \nu(dz)$, where $\nu(dz)$ admits the density $\nu(dz) = (20 e^{-z}/z) \mathds{1}_{z>0}\, dz$. 
Note that $\nu(dz)$, and hence $\lambda(dt,dz)$, is an infinite measure, with a singularity around zero.
Accordingly, the process $X_t$ has infinitely many small, positive jumps. 
This is possible because the singularity of $\nu(dz)$ is small enough for the jumps to be summable.
 
\section{Nonstationary jump processes}\label{nonstationary-jump-processes}

Defining a stochastic process \(X_t\) via \(\eqref{eqn:jump-integral}\) allows us to relate various properties of \(X_t\) to the Poisson random measure \(N(ds,dz)\) and the corresponding intensity measure \(\lambda(ds,dz)\). 
For example, for two Borel sets \(A,B\subset [0,\infty)\times\R^d\) such that \(A\cap B=\emptyset\), the random variables \(N(A)\) and \(N(B)\) are stochastically independent.
As a consequence, the increments \(X_{t+s_1+s_2}-X_{t+s_1}\) and \(X_{t+s_1}-X_t\) are independent as well, just as for a Lévy process.
For \(X_t\) to be a Lévy process, the increments furthermore need to be stationary, i.e.\ the distribution of the random vectors $X_{t+h}-X_t$ should not depend on $t$, but only on $h>0$.
This property holds if the intensity measure \(\lambda(ds,dz)\) is homogeneous in time, that is, we may write it as a product measure \(\lambda(ds,dz) = ds \, \nu(dz)\), where \(ds\) denotes the Lebesgue measure on \([0,\infty)\), and \(\nu(dz)\) is a measure on \(\R^d\). 
In this stationary case, \(\nu(dz)\) is called the Lévy measure of the Lévy process \(X_t\). 
One readily finds that for any Borel set \(C\subset\R^d\), we have \(\nu(C)=\lambda([0,1]\times C) = \mathbb{E}( \sum_{t\in[0,1]} \mathds{1}(\Delta X_t \in C) )\), i.e.\ \(\nu(C)\) is the expected number of jumps of size \(\Delta X_t \in C\) during the unit time interval.

If the intensity measure \(\lambda(ds,dz)\) is not of product form, then the process \(X_t\) is in general nonstationary. 
This nonstationarity is deterministic by nature, since the stochastic behavior of \(X_t\) in the interval \([1,2]\), say, is fully determined by the non-random measure \(\lambda(ds,dz)\). 
However, from a modeling perspective, it is of interest to let the random trajectory of \(X_t, t\in[0,1]\) affect the behavior of \(X_t\) on the interval \([1,2]\). 
For example, the jump behavior of \(X_t\) at time \(t\) could depend on the value \(X_t\), giving rise to a special kind of Markov process. 
If \(X_t\) describes the location of a particle, this would mean that the jump behavior is spatially inhomogeneous. 
This state-dependent behavior gives rise to a stochastic differential equation. 
An established method to formalize this is to maintain a deterministic intensity measure \(\lambda\), while defining the process \(X_t\) differently as
\begin{align}
	X_t = \int_0^t \int_{\R^d} c(s,z, X_{t-})\, N(ds,dz). \label{eqn:def-sdejump}
\end{align}
It is important to use the left-hand limit \(X_{t-}\), because the value \(X_t\) may already be affected by a jump at time \(t\). 
Furthermore, for the integral\(\eqref{eqn:def-sdejump}\) to be well-defined as a countable sum, the function \(c(s,z,x)\) needs to be such that \(\int_{0}^t \int_{\R^d} \mathbb{E}\|c(s,z,X_{t-})\| \, \lambda(ds,dz)<\infty\).
Verifying this condition is in general non-trivial and amounts to establishing existence of solutions of stochastic differential equations. 
Simple conditions can be formulated in terms of Lipschitz-continuity of the mapping \(x\mapsto c(s,z,x)\), see Theorem \ref{thm:jd-existence} below.
In the sequel, we assume that the function \(c(s,z,x)\) is such that \(\eqref{eqn:def-sdejump}\) is well-defined.

In \eqref{eqn:def-sdejump}, the variable $z$ is called the mark of the Poisson process, which is transformed to a jump of size $c(s,z, X_{t-})$. 
Here, we choose the mark space $\R^d$, i.e.\ $z\in\R^d$, but the theory for Poisson integrals is formulated for more general mark spaces $\mathcal{E}$. 

For many purposes, it is most suitable to choose a homogeneous basic intensity measure, i.e.\ \(\lambda(ds,dz) = ds \otimes \nu(dz)\) for some measure \(\nu\), and to introduce nonstationarity via \eqref{eqn:def-sdejump}. 
In particular, if \(X_t\) is given by \eqref{eqn:def-sdejump}, we may describe the instantaneous jump behavior by the measure \(\nu_t(dz)\), which is the push-forward measure of \(\nu(dz)\) under the mapping \(z\mapsto c(t,z,X_{t-})\). 
That is, for any Borel set \(C\subset \R^d\),
\begin{align}
	\nu_t(A) = \nu\left( \{ z\in\R^d: c(t,z,X_{t-})\in A \} \right). \label{eqn:push-forward}
\end{align}
In fact, if we change the underlying intensity measure \(\nu(dz)\) and the mapping \(c(s,z,x)\) such that \(\nu_t(dz)\) remains unchanged, then the corresponding stochastic process \(X_t\) is the same, distributionally. 
Hence, the measure \(\nu_t(dz)\) fully describes the jump behavior of \(X_t\). Note that \(\nu_t(dz)\) depends implicitly on \(X_{t-}\), i.e. \(\nu_t(dz)\) is a random measure. 
Instead of modeling the mapping \(X_{t-}\mapsto c(t,z,X_{t-})\), it is often more intuitive to model the mapping \(X_{t-}\mapsto \nu_t(dz;X_{t-})\) in order to specify the instantaneous jump behavior. 
The measure \(\nu_t(dz)\) is called the instantaneous intensity measure, or spot intensity measure in the mathematical finance literature. 
The measure should be specified such that \(\int_{\R^d} \|z\| \nu_t(dz) < \infty \) almost surely. 
This condition can be weakened to \(\int_{\R^d} \min(1, \|z\|^2)\, \nu_t(dz)<\infty\) using the more general \(L_2(P)\) construction of jump integrals.
However, we do not need this generality in the present paper.

\section{Jump Diffusions}\label{jump-diffusions}

\begin{figure}[tb]
	\centering
	\includegraphics[width=0.85\textwidth]{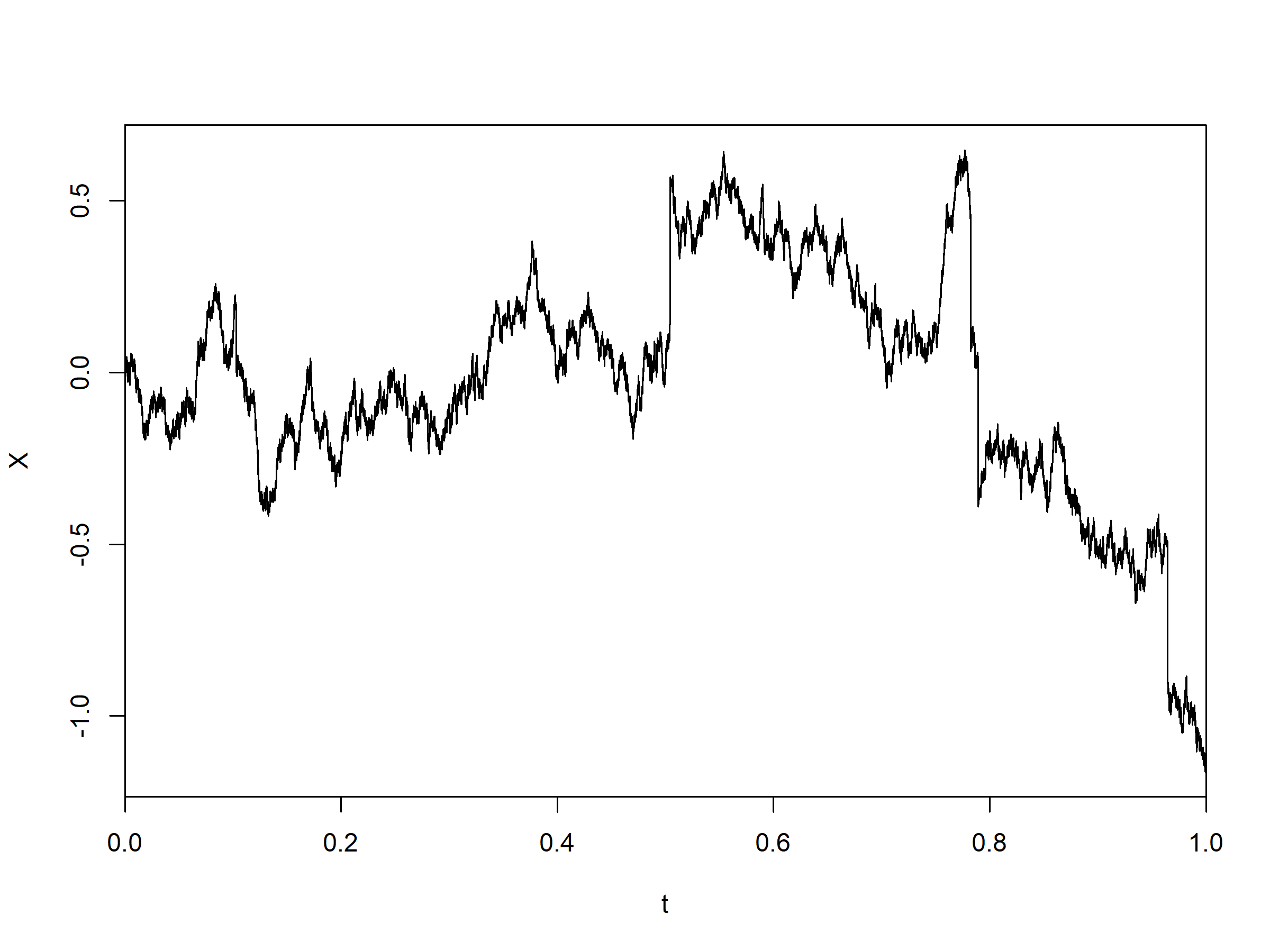}
	\caption{Sample path of a Lévy process, as a special instance of a jump diffusion. In this case $X_t=B_t + J_t$ for a standard Brownian motion $B_t$, and $J_t$ as in Figure \ref{fig:compount-poisson}.}
	\label{fig:JD}
\end{figure}

One may also construct processes which combine a dynamic jump behavior as described above, with a continuous Gaussian diffusion between jumps, see Figure \ref{fig:JD}. 
Stochastic processes of this type are called jump diffusions. 
If the process has infinitely many jumps, the notion of time ``between jumps'' is not sensible, but the jump diffusion may still be defined via stochastic calculus.

A jump diffusion is a stochastic process \(X_t\) which satisfies a stochastic differential equation consisting of
\begin{itemize}
	\item a drift term \(\mu(t, X_{t-})\),
	\item a diffusion term with diffusion matrix \(\Sigma(t,X_{t-})\), and
	\item a jump term with instantaneous intensity measure
	\(\nu_t(dz; X_{t-})\).
\end{itemize}
In particular, the process needs to satisfy the stochastic integral equation
\begin{align}
	X_t = X_0 + \int_0^t \mu(s,X_{s-})\, ds + \int_0^t \Sigma^{\frac{1}{2}}(s,X_{s-}) \, dW_s + \int_0^t \int_{\R^d} c(s,z, X_{s-})\, N(ds,dz). \label{eqn:jump-diffusion}
\end{align}
Here, the mapping \(c(s,z,x)\) and the intensity measure \(\lambda(ds,dz)=ds \, \nu(dz)\) need to be such that the measure \(\nu_t(dz; X_{t-})\) is as specified in \(\eqref{eqn:push-forward}\).
The process \(W_s\) is a standard \(d\)-dimensional Brownian motion also known as Wiener process, and \(\int_0^t (\cdot) \, dW_s\) denotes the Itô integral.
Most mathematical results on jump diffusions employ weaker assumptions on the function $c(s,z,x)$ of the form $\int_0^t \int_{\R^d} \E \min(1, \|c(s,z, X_{s-}\|^2)\, \nu(dz)\, ds < \infty$.
To make sense of the Poisson integral in this more general setting, one studies the equation
\begin{align}
    \begin{split}
	X_t &= X_0 + \int_0^t \mu(s,X_{s-})\, ds + \int_0^t \Sigma^{\frac{1}{2}}(s,X_{s-}) \, dW_s \\
	&\qquad + \int_0^t \int_{\R^d} c(s,z, X_{s-})\, (N-\lambda)(ds,dz),
	\end{split}\label{eqn:jump-diffusion-compensated}
\end{align}
where $(N-\lambda)$ is the so-called \emph{compensated Poisson random measure}.
Existence and uniqueness of the solution of this SDE may be derived under suitable Lipschitz conditions.
If the transformation $c(s,z,x)$ is such that $\tilde{\mu}(s,x) \int_{\R^d} c(s,z,x)\, \nu(dz)<\infty$ is sufficiently regular, then we may write \eqref{eqn:jump-diffusion} in the form \eqref{eqn:jump-diffusion-compensated} with drift $\bar{\mu} = \mu - \tilde{\mu}$.
Hence, one may easily deduce results for equation \eqref{eqn:jump-diffusion-compensated} from the more general results in the literature.

The following theorem gives conditions for the existence of solutions to \eqref{eqn:jump-diffusion-compensated} in terms of Lipschitz-continuity. 
Denote $a\vee b= \max(a,b)$.

\begin{theorem}[\cite{kunita2019}, Theorem\ 3.3.1 \& Lemma 3.3.2]\label{thm:jd-existence}
	Suppose that for \(p\geq 2\), there exists some \(C>0\) such that the coefficients of \(\eqref{eqn:jump-diffusion-compensated}\) are bounded and Lipschitz such that
	\begin{align*}
		\| \mu(t,0)\| + \|\Sigma^{\frac{1}{2}}(t,0)\| + \int_{\R^d} \|c(s,z,0)\|^p (|z|^{p-2}\vee 1) \nu(dz) \leq C, 
	\end{align*}
	and
	\begin{align*}
		&\|\mu(t,x)-\mu(t,y)\| + \|\Sigma^{\frac{1}{2}}(t,x) - \Sigma^{\frac{1}{2}}(t,y)\| \\
		&\qquad+ \left(\int_{\R^d} \|c(s,z,x) - c(s,z,y)\|^p (|z|^{p-2}\vee 1) \nu(dz)\right)^{\frac{1}{p}}  \leq C \|x-y\|.
	\end{align*}
	Then equation \(\eqref{eqn:jump-diffusion-compensated}\) admits a unique solution.
	If \(X_0\in L_p(P)\), then \(X_t\in L_p(P)\) for all \(t\geq 0\).
\end{theorem}

Now denote the (marginal) distribution of \(X_t\), for any \(t\geq 0\), by \(P_t\). 
That is, the initial value follows the distribution \(X_0\sim P_0\), and \(P_t\) describes the evolution of this distribution. 
For a jump diffusion as above, we may employ Dynkin's formula \citep[Thm.\ 1.22] {oksendal2007} to find that\footnote{The additional term in the integral in \citet{oksendal2007} occurs because they consider equation \eqref{eqn:jump-diffusion-compensated} instead of \eqref{eqn:jump-diffusion}.}, for any \(f\in C^2_0\),
\begin{align}
	\int f\, dP_t &= \mathbb{E} f(X_t) \overset{!}{=} \mathbb{E} f(X_0) + \int_0^t \mathbb{E} [\mathcal{A}_sf(X_s)]\, ds = \int f\, dP_0 + \int_0^t \int \mathcal{A}_sf\, dP_s, \label{eqn:dynkin} \\
	(\mathcal{A}_tf)(x) &= - \sum_{i=1}^d \partial_i f(x) \cdot  \mu(t,x)_i - \frac{1}{2} \sum_{i,j=1}^d \partial_{ij}f(x)\cdot \Sigma(t,x)_{ij} \nonumber \\
	&\qquad + \int_{\R^d} [f(x+c(t,z,x))-f(x)]\, \nu(dz). \nonumber
\end{align}
The operator \(\mathcal{A}_t\) is called the generator of the Markov process \(X_t\), and gives rise to a functional analytic perspective on stochastic processes. 
See \cite{ethier2009markov} for more details.

Dynkin's formula may be extended to a broader class of functions \(f\)
if we impose stronger assumptions on the process \(X_t\).

\begin{proposition}\label{prop:dynkin-Lp}
	Let the conditions of Theorem \ref{thm:jd-existence} hold for all \(p\in[1,2+q]\), for some \(q>0\). 
	Then Dynkin's formula \(\eqref{eqn:dynkin}\) holds for any function \(f\in C^2(\R^d)\) such	that 
	\(|f(x)| + \|Df(x)\| + \|D^2f(x)\| \leq C(1+\|x\|)^q\). 
	Moreover, there exists some \(\tilde{C}\) such that
	\(|\mathcal{A}_t f(x)|\leq \tilde{C}(1+\|x\|)^{q+2}\) 
	for all \(t\geq 0\).
\end{proposition}
\begin{proof}[Proof of Proposition \ref{prop:dynkin-Lp}]
	Any such function \(f\) may be approximated by a sequence \(f_n\in C_0^2(\R^d)\) such that
	\(|D^kf_n(x)| \leq \tilde{C}(1+\|x\|)^q\) 
	and \(D^kf_n(x)\to D^kf(x)\) pointwise, for \(k=0,1,2\). 
	For example, we may set \(f_n = f\cdot \varphi_n\), where \(\varphi_n\) is a sequence of \(C^\infty_0\) indicators on \([-n,n]\) with bounded derivatives up to order \(2\). 
	Via dominated convergence, we find that
	\(\int f_n \, dP_t \to \int f \, dP_t\), and
	\(\int \partial_i f_n(x)\mu(t,x)\, dP_t(x) \to \int \partial_i f(x)\mu(t,x)\, dP_t(x)\), and
	\(\int \partial_{ij}f_n(x) \Sigma(t, x)_{ij}\, dP_t(x) \to \int \partial_{ij}f_n(x) \Sigma(t, x)_{ij}\, dP_t(x)\).
	
	It remains to study the jump term, i.e. the integral operator. 
	To this end, we employ the mean value theorem, i.e.\ for some \(\tilde{x}(t,z)\) between \(x\) and \(x+c(t,z,x)\),
	\begin{align*}
		& \int_{\R^d} \left|f(x+c(t,z,x))-f(x)\right|\, \nu(dz) \\
		&\leq \int_{\R^d} \|Df(\tilde{x}(t,z))\|\cdot \|c(t,z,x)\|\, \nu(dz) \\
		&\leq \int_{\R^d} C (1+\|x\|+\|c(t,z,x)\|)^q\cdot \|c(t,z,x)\|\, \nu(dz) \\
		&\leq \tilde{C}\int_{\R^d} \left[ (1+\|x\|)^{q} \|c(t,z,x)\| + \|c(t,z,x)\|^{q+1}\right]\, \nu(dz) \\
		&\leq \tilde{C}(1+\|x\|)^{q} \int_{\R^d} \left[  \|c(t,z,0)\| + \|c(t,z,x)-c(t,z,0)\| \right] \, \nu(dz) \\
		&\quad + \tilde{C} \int_{\R^d} \left[ \|c(t,z,0)\|^{q+1} + \|c(t,z,x)-c(t,z,0)\|^{q+1}\right]\, \nu(dz) \\
		&\overset{(*)}{\leq} \tilde{C}(1+\|x\|)^q(1+\|x\|) \; + \; \tilde{C} (1+\|x\|^{q+1}) \\
		&\leq \tilde{C} (1+\|x\|^{q+1}).
	\end{align*}
	
	At step \((*)\), we used the conditions of Theorem \ref{thm:jd-existence} for \(p=1\) and \(p=q+1\). 
	Also, the factor \(\tilde{C}\) may vary from line to line.
	
	Note that the upper bound \((1+\|x\|^{q+1})\) is integrable with respect to \(P_t\). 
	Hence,	
	\(|f(x+c(t,z,x))-f(x)| \leq C (1+\|x\| + \|c(t,z,x)\| )^q \cdot \|c(t,z,x)\|\)
	is a majorant which is integrable w.r.t.\
	\(P_t(dx)\otimes \nu(dz)\otimes dt\). 
	Replacing \(C\) by some \(\tilde{C}\), this is also a majorant for	\(|f_n(x+c(t,z,x)) - f_n(x)|\). 
	The dominated convergence theorem yields
	 \begin{align*}
	 	&\int_0^t \int\int_{\R^d} \left[f_n(x+c(s,z,x))-f_n(x)\right]\, \nu(dz)\, P_s(dx)\, ds \\
	 	\overset{n\to\infty}{\longrightarrow} & \int_0^t \int\int_{\R^d} \left[f(x+c(s,z,x))-f(x)\right]\, \nu(dz)\, P_t(dx)\, ds.
	 \end{align*}
	Hence, we may pass to the limit \(f_n\to f\) on both sides of
	\(\eqref{eqn:dynkin}\).
\end{proof}

Differentiating equation \eqref{eqn:dynkin} yields the identity
\begin{align}
	\frac{d}{dt} \int f\, dP_t = \int \mathcal{A}_t f\, dP_t. \label{eqn:dynkin-local}
\end{align}
Thus, we obtain an evolution equation for the nonlinear moments of the distribution \(P_t\).
If the distribution \(P_t\) admits a density, i.e.\ \(P_t(dx)=p_t(x)\, dx\), then equation \eqref{eqn:dynkin} may be rewritten as
\begin{align*}
	\langle f, p_t\rangle 
	&= \langle f, p_0\rangle + \int_0^t \langle \mathcal{A}_sf, p_s\rangle \, ds \\
	&= \langle f, p_0\rangle +  \langle f, \smallint_0^t \mathcal{A}_s^* p_s\rangle \, ds,
\end{align*}
where \(\langle f,g\rangle = \int f(x)g(x)\, dx\) denotes the scalar product in \(L_2(\R^d)\), and \(\mathcal{A}_t^*\) is the formal adjoint operator of \(\mathcal{A}_t\). 
Since Dynkin's formula holds in particular for all \(f\in C_0^\infty\), the variational formulation yields that \(p_t = p_0 + \int_0^t \mathcal{A}_s^* p_s\), or 
\begin{align}
	\frac{d}{dt}p_t = \mathcal{A}_t^* p_t. \label{eqn:forward}
\end{align}

Equation \eqref{eqn:forward} is known as the \emph{Kolmogorov forward equation} in the literature. 
It should be pointed out that the rigorous derivation of \(\eqref{eqn:forward}\) raises several difficulties. 
In particular, the density \(p_t\) needs to exist and be sufficiently regular such that \(\mathcal{A}_t^*p_t\) is well-defined. 
Furthermore, the adjoint operator \(\mathcal{A}^*_t\) is in general hard to determine due to the jump integral operator. 

In the context of this article, we may exploit some additional structure.
In particular, the state of a particle is described by a location vector and a velocity vector.
On the one hand, the instantaneous jump intensity depends on the location of the particle, but not on its velocity.
On the other hand, our proposed model only includes jumps in velocity, but not in location. 
That is, we may represent the model by a transfer function $c(t,z,x)$ such that 
\begin{align}
	c(t,z,x) = c(t,z,x+c(t,z,x)). \label{eqn:jump-compatibile}
\end{align}

\begin{proposition}\label{prop:jump-compatible}
	Suppose that $x\mapsto c(t,z,x)$ is continuously differentiable for all $t,z$.
	If \eqref{eqn:jump-compatibile} holds, then the formal adjoint operator $\mathcal{A}^*_t$ is well defined for any $g\in C^2_0$, and is given by
	\begin{align}
		\begin{split}
		\mathcal{A}^*_t g (x) 
		&= \sum_{i=1}^d \partial_i \left[g(x) \mu(t,x)_i\right] - \frac{1}{2} \sum_{i,j=1}^d \partial_{ij} \left[ g(x) \Sigma(t,x)_{ij} \right] \\
		&\qquad + \int_{\R^d} \left[g(x-c(t,z,x))-g(x)\right]\, \nu(dz)
		\end{split}\label{eqn:adjoint}
	\end{align}
\end{proposition}

We call $\mathcal{A}^*_t$ in Proposition \ref{prop:jump-compatible} the \emph{formal} adjoint because we discuss neither the domain of $\mathcal{A}_t$ nor the domain of $\mathcal{A}_t^*$. 

\begin{proof}[Proof of Proposition \ref{prop:jump-compatible}]
	The adjoints of the drift and the diffusion part may be derived via partial integration.
	Hence, we study the jump term. 
	If $\nu(\R^d)<\infty$, we have
	\begin{align*}
		&\int g(x) \int \left[f(x+c(t,z,x)) - f(x)\right]\, \nu(dz)\, dx \\
		&= \int \int g(x) f(x+c(t,z,x)) \,dx\, \nu(dz)  - \int g(x)f(x) \int \, \nu(dz)\, dx.
	\end{align*}
	Now, observe that the equation $y=x+c(t,z,x)$ has the unique solution $x=y-c(t,z,y)$. 
	A substitution yields
	\begin{align*}
		&\int \int g(x) f(x+c(t,z,x)) \,dx\, \nu(dz) \\
		&= \int \int g(x) f(x+c(t,z,x+c(t,z,x))) \,dx\, \nu(dz) \\
		&= \int \int g(y-c(t,z,y)) f(y-c(t,z,y)+c(t,z,y)) \left| \det\left( I-D_y c(t,z,y) \right) \right| \,dx\, \nu(dz) \\
		&= \int \int g(y-c(t,z,y)) f(y) \left| \det\left( I-D_y c(t,z,y) \right) \right| \,dx\, \nu(dz).
	\end{align*}
	Lemma \ref{lem:c-determinant} below shows that $\left| \det\left( I-D_y c(t,z,y) \right) \right|=1$.
	Hence,
	\begin{align*}
		\int g(x) \int \left[f(x+c(t,z,x)) - f(x)\right]\, \nu(dz)\, dx
		&= \int\int \left[g(y-c(t,z,y)) - g(y)\right] f(y)\, \nu(dz)\, dy.
	\end{align*}
	This identity for the case $\nu(\R^d)<\infty$, and may be extended to the general case by a limiting procedure, using that $\int \|c(t,z,y)\|\, \nu(dy)<\infty$.
\end{proof}

\begin{lemma}\label{lem:c-determinant}
	Let $h:\R^d\to\R^d$, $h\in C^1$, such that $h(x)=h(x+h(x))$ for all $x\in\R^d$. 
	Then $|\det(I-D_xh(x))|=1$ for all $x\in\R^d$.
\end{lemma}
\begin{proof}[Proof of Lemma \ref{lem:c-determinant}]
	Let $f(x)=x+h(x)$, and determine its inverse as
	\begin{align*}
		&y\overset{!}{=}x+h(x) = x+h(x+h(x)) = x+h(y), \\
		\leadsto& x=y-h(y),
	\end{align*}
	i.e.\ $f^{-1}(y) = y-h(y)$ for all $y\in\R^d$. 
	fix some $x\in\R^d$, and denote $A=D_x h(x)$. 
	Then $D_xf(x) = I+A$ and $D_xf^{-1}(x)=I-A$. 
	On the other hand, $D_xf^{-1}(x) = [D_xf(x)]^{-1}$.
	Hence, we may conclude that 
	\begin{align}
		I-A = (I+A)^{-1}. \label{eqn:A-det}
	\end{align}
	We will show that \eqref{eqn:A-det} implies $|\det(I+A)| = |\det(I-A)| =1$.
	
	A consequence of \eqref{eqn:A-det} is that $I=(I-A)(I+A)=I-AA$, so that $AA=0$.
	Thus, $(I+A)^2 = I + 2A +A^2 = I+2A$, and via induction
	\begin{align*}
		(I+A)^n = I+nA,\qquad n\in\N.
	\end{align*}
	Using the mulitiplicativity of the determinant, $\frac{|\det(I+A)|^n}{n} = |\det(A + I/n)|$. 
	Now let $n\to\infty$ and use the continuity of the determinant, to find that
	\begin{align*}
		\lim_{n\to\infty} \frac{|\det(I+A)|^n}{n} = \det(A)=0.
	\end{align*}
	This implies $|\det(I+A)|\leq 1$, since otherwise the left hand side diverges.
	By the symmetry of \eqref{eqn:A-det}, we may repeat the argument for $-A$ to find that $|\det(I-A)|\leq 1$. 
	Since $\det(I-A) = \det((I+A)^{-1}) = 1/\det(I+A)$, we conclude that $|\det(I+A)|=1$.
\end{proof}




\bibliography{levyBoltzmann,levyBoltzmann2}
\bibliographystyle{apalike}